\documentclass[letterpaper,twocolumn,10pt]{article}
\usepackage{usenix-2020-09}

\usepackage{amsmath,amssymb,amsfonts}
\usepackage{graphicx}
\usepackage{booktabs}
\usepackage{algorithm}
\usepackage{algpseudocode}
\usepackage{listings}
\usepackage{multirow}
\usepackage{subcaption}

\begin{document}

\date{}

\title{\Large \bf Retrieval Pivot Attacks in Hybrid RAG:\\
Measuring and Mitigating Amplified Leakage from\\
Vector Seeds to Graph Expansion}

\author{
{\rm Scott Thornton}\\
scott@perfecxion.ai
}

\maketitle

\begin{abstract}
Hybrid Retrieval-Augmented Generation (RAG) pipelines increasingly
combine vector similarity search with knowledge graph expansion to
support multi-hop reasoning. We show this composition introduces a
distinct security failure mode: a semantically retrieved ``seed'' chunk
can pivot via entity linking into sensitive graph neighborhoods,
causing data leakage that does not exist in vector-only retrieval. We
formalize this risk as \textbf{Retrieval Pivot Risk (RPR)} and define
companion metrics \textbf{Leakage@k}, \textbf{Amplification Factor
(AF)}, and \textbf{Pivot Depth (PD)} to quantify leakage probability,
magnitude, amplification over vector baselines, and structural distance
to the first unauthorized node. We further present a taxonomy of four
\textbf{Retrieval Pivot Attacks} that exploit the vector-to-graph
boundary with injection budgets as small as 10--20 chunks, and we
demonstrate that adversarial injection is not required: naturally
shared entities (e.g., vendors, infrastructure, compliance standards)
create cross-tenant pivot paths organically. In a synthetic
multi-tenant enterprise corpus (1{,}000 documents; 2{,}785 graph
nodes; 15{,}514 edges) evaluated with 500 queries, the undefended
hybrid pipeline exhibits $\text{RPR} \approx 0.95$ and
$\text{AF}(\epsilon) \approx 160\text{--}194\times$ relative to
vector-only retrieval, with leakage occurring at $\text{PD} = 2$ hops
as a structural consequence of the bipartite chunk--entity topology.
We validate these findings on two real-world corpora: the Enron email
corpus (50{,}000 emails; $\text{RPR} = 0.695$) and SEC EDGAR 10-K
filings (887 sections across 20 companies; $\text{RPR} = 0.085$).
RPR scales with entity connectivity density, but the structural
invariant ($\text{PD} = 2$) persists across all three corpora.
We propose five layered defenses and find that a single placement
fix---\textbf{per-hop authorization at the graph expansion
boundary}---eliminates leakage in our evaluation
($\text{RPR} \to 0.0$) across all three corpora, all queries, and
all attack variants with negligible latency overhead, indicating the
vulnerability is primarily a boundary enforcement problem rather than
a defense complexity problem.
\end{abstract}

\section{Introduction}
\label{sec:introduction}

Enterprise adoption of Retrieval-Augmented Generation (RAG) has
accelerated rapidly: 30--60\% of enterprise AI use cases now rely on
RAG architectures~\cite{saferag2025}, and the vector database market
reached \$1.73 billion in 2024~\cite{pineconerac2024}. To improve
multi-hop reasoning over complex organizational knowledge,
practitioners increasingly deploy \emph{hybrid} RAG pipelines that
combine vector similarity search with knowledge graph
expansion~\cite{hybridrag2024, microsoftgraphrag2024}---a pattern
supported by production frameworks including LangChain, LlamaIndex,
and Microsoft GraphRAG. These systems
retrieve an initial set of text chunks via embedding similarity, then
\emph{pivot} through entity mentions into a knowledge graph to gather
structurally related context before passing the assembled results to a
large language model (LLM). We call the transition between vector
retrieval and graph expansion the \textbf{pivot boundary}: the
architectural point where a retrieved seed becomes a graph traversal
capability.

\begin{figure}[t]
\centering
\includegraphics[width=0.65\columnwidth]{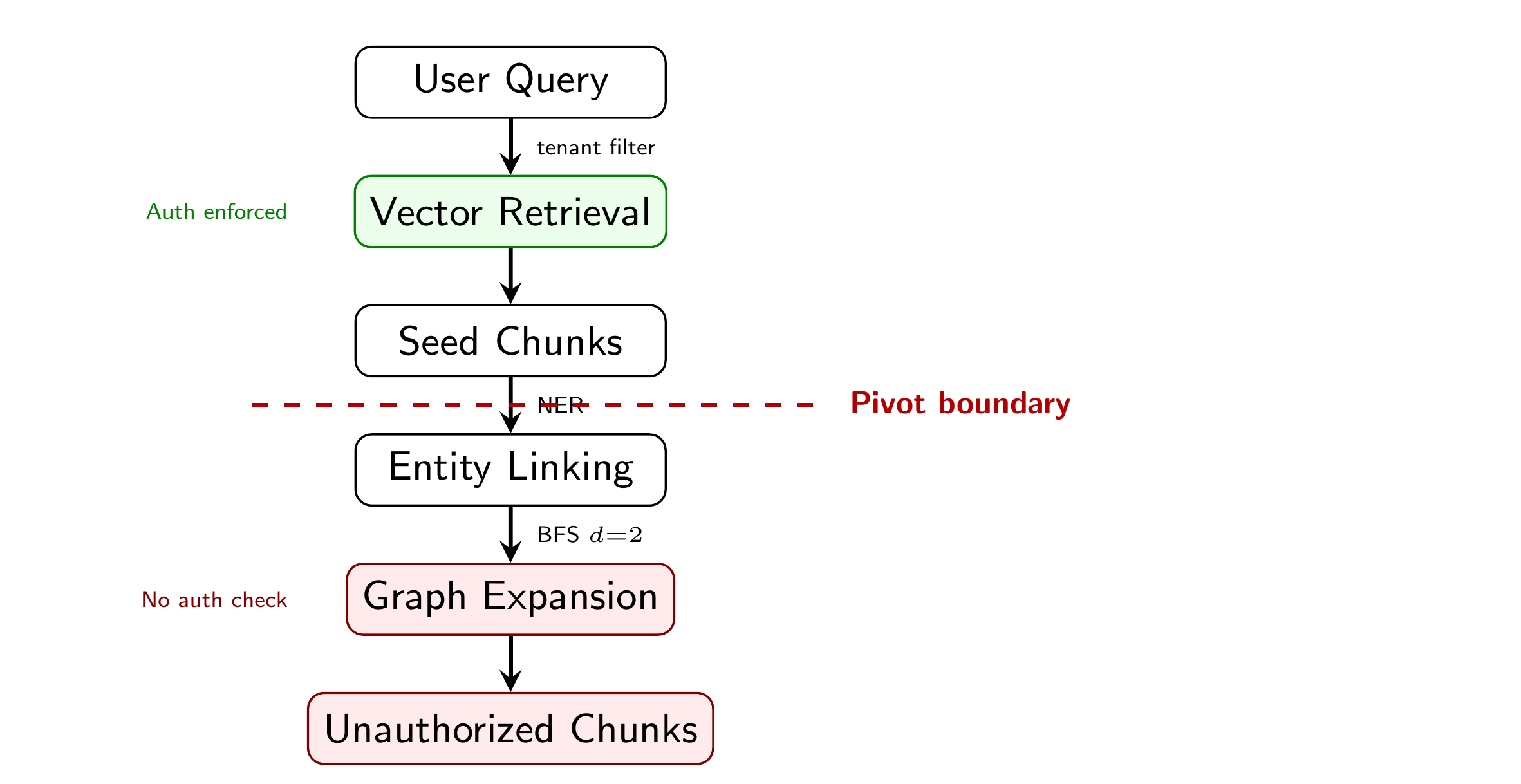}
\caption{The hybrid RAG pipeline. Authorization is enforced at vector
retrieval but absent during graph expansion, creating the pivot
boundary where cross-tenant leakage occurs.}
\label{fig:pipeline}
\end{figure}

This boundary is a \emph{boundary placement} bug analogous to
classical confused-deputy vulnerabilities~\cite{hardy1988confused}: the
graph traversal engine---a component with access to the entire
knowledge graph---executes expansion on behalf of a tenant-restricted
caller, inheriting the caller's seed but not the caller's access
constraints. The vector store enforces tenant policy \emph{before}
retrieval. The graph expansion must enforce it \emph{again} after
entity linking---otherwise an authorized seed chunk becomes an
uncontrolled entry point into the knowledge graph. Consider an analyst at an engineering firm who queries
about Kubernetes cluster configurations. Vector retrieval returns
authorized engineering documents. But those documents mention shared
entities (``CloudCorp,'' ``auth-service'') that also appear in the
knowledge graph connected to confidential HR salary records, restricted
security credentials, and financial audit data belonging to other
tenants. A 2-hop graph expansion traverses through these shared entities
into unauthorized neighborhoods, silently injecting sensitive
cross-tenant data into the analyst's context window.
Figure~\ref{fig:pipeline} illustrates this pipeline and the pivot
boundary where leakage occurs.

Prior work falls into two separate buckets that do not address this
boundary. \textbf{(1) Vector-only attacks}: PoisonedRAG achieves
90\% attack success rate by injecting 5 malicious texts into vector
stores~\cite{zou2025poisonedrag}; CorruptRAG~\cite{corruptrag2025} and
CtrlRAG~\cite{ctrlrag2025} extend these results.
\textbf{(2) Graph-only attacks}: GRAGPoison demonstrates 98\% success
via relation-centric poisoning within
GraphRAG~\cite{gragpoison2026}; TKPA~\cite{fewwords2025} and
RAG-Safety~\cite{ragsafety2025} target graph-side integrity.
\textbf{(3) The hybrid boundary}---how vector outputs become graph
seeds---has not been systematically measured or formalized. The
OWASP LLM Top 10 (2025) introduced LLM08 (Vector and Embedding
Weaknesses) but does not address graph
components~\cite{owasptop10llm2025}. MITRE ATLAS treats RAG as a
monolithic system~\cite{mitreatlas2025}. The SoK on RAG privacy
explicitly identifies hybrid RAG security as an open
problem~\cite{sokprivacy2026}.

This paper makes three contributions:

\begin{enumerate}
\item \textbf{Composition vulnerability + metrics.} We formalize
  \emph{Retrieval Pivot Risk (RPR)} and companion metrics (AF, PD,
  Leakage@k, severity-weighted leakage) that quantify a compound
  attack surface emerging from composing two individually secure
  retrieval modalities. In our bipartite chunk--entity graphs,
  all leakage occurs at PD\,$=$\,2 hops---a structural property of
  bipartite entity-link topologies (knowledge graphs with richer
  entity-to-entity relations may exhibit deeper pivot paths)
  (\S\ref{sec:metrics}).

\item \textbf{Cross-dataset validation + organic leakage.} We
  demonstrate the vulnerability on three corpora: a synthetic enterprise
  corpus (1{,}000 documents, 4 tenants; RPR\,$=$\,0.95), the Enron email
  corpus (50{,}000 emails, 5 departments; RPR\,$=$\,0.70), and SEC EDGAR
  10-K filings (887 sections, 4 sectors; RPR\,$=$\,0.09)---all without
  adversarial injection. We present four
  Retrieval Pivot Attacks (A1--A4) exploiting the pivot boundary with
  injection budgets of 10--20 chunks
  (\S\ref{sec:attacks}, \S\ref{sec:results}).

\item \textbf{Defense placement analysis.} We evaluate five layered
  defenses (D1--D5) and show that authorization must be re-checked at
  the graph expansion boundary. D1 (per-hop authorization) alone
  eliminates all measured leakage ($\text{RPR} \to 0.0$) across both
  corpora, all queries, all attack variants, and metadata mislabel
  rates up to 5\%. D2--D5 serve as utility optimizers and
  defense-in-depth layers
  (\S\ref{sec:mitigations}, \S\ref{sec:results}).
\end{enumerate}

The complete codebase, data generators, attack implementations,
defense suite, and experimental results are available at
\url{https://github.com/scthornton/hybrid-rag-pivot-attacks}.

\section{Background}
\label{sec:background}

\subsection{Vector Retrieval in RAG}

Standard RAG systems encode documents as dense vector embeddings using
models such as all-MiniLM-L6-v2~\cite{corpuspoisoning2023} and
retrieve the top-$k$ chunks by cosine similarity to the query
embedding. In multi-tenant deployments, vector stores apply metadata
prefilters (e.g., tenant ID) before similarity search, ensuring that
retrieval respects organizational boundaries. This prefiltering makes
vector-only RAG robust against cross-tenant leakage: our experiments
confirm $\text{RPR} = 0.0$ for vector-only pipelines across all
evaluation queries on all three corpora
(\S\ref{sec:results}).

\subsection{Knowledge Graph RAG}

GraphRAG~\cite{microsoftgraphrag2024} and related approaches
construct knowledge graphs from document corpora via named entity
recognition (NER) and relation extraction, then leverage graph
structure for multi-hop reasoning. Systems like
AgCyRAG~\cite{agcyrag2025} use LLM-driven graph traversal for
complex queries. Graph-based retrieval provides superior
comprehensiveness on multi-hop queries---86\% versus 57\% for vector
RAG~\cite{hybridrag2024}---but introduces graph-specific attack
surfaces: GRAGPoison~\cite{gragpoison2026} achieves 98\% attack
success through relation-centric poisoning, and TKPA~\cite{fewwords2025}
drops QA accuracy from 95\% to 50\% by modifying just 0.06\% of
corpus text.

\subsection{Hybrid RAG and the Pivot Boundary}

Hybrid RAG combines both retrieval modalities in a pipeline:

\begin{enumerate}
\item \textbf{Vector retrieval:} Query embedding $\rightarrow$ cosine
  similarity $\rightarrow$ top-$k$ seed chunks $S(q)$.
\item \textbf{Graph seeding:} Map each seed chunk to its corresponding
  node in the knowledge graph (via chunk ID).
\item \textbf{Graph expansion:} BFS traversal from seed chunk nodes
  to depth $d$, following MENTIONS edges to entity nodes and onward
  to structurally related chunks.
\item \textbf{Context merge:} Combine vector-retrieved chunks with
  graph-expanded context for LLM generation.
\end{enumerate}

We define the \textbf{pivot boundary} as the transition between steps
1--2 and steps 3--4: the point where vector retrieval results become
graph traversal seeds. This boundary is the attack surface we study.
Vector-side prefilters operate \emph{before} the pivot; graph
expansion operates \emph{after} it. If graph expansion does not
independently enforce access controls, any entity mentioned in an
authorized seed chunk becomes an uncontrolled entry point into the
knowledge graph.

\begin{figure}[t]
\centering
\includegraphics[width=\columnwidth]{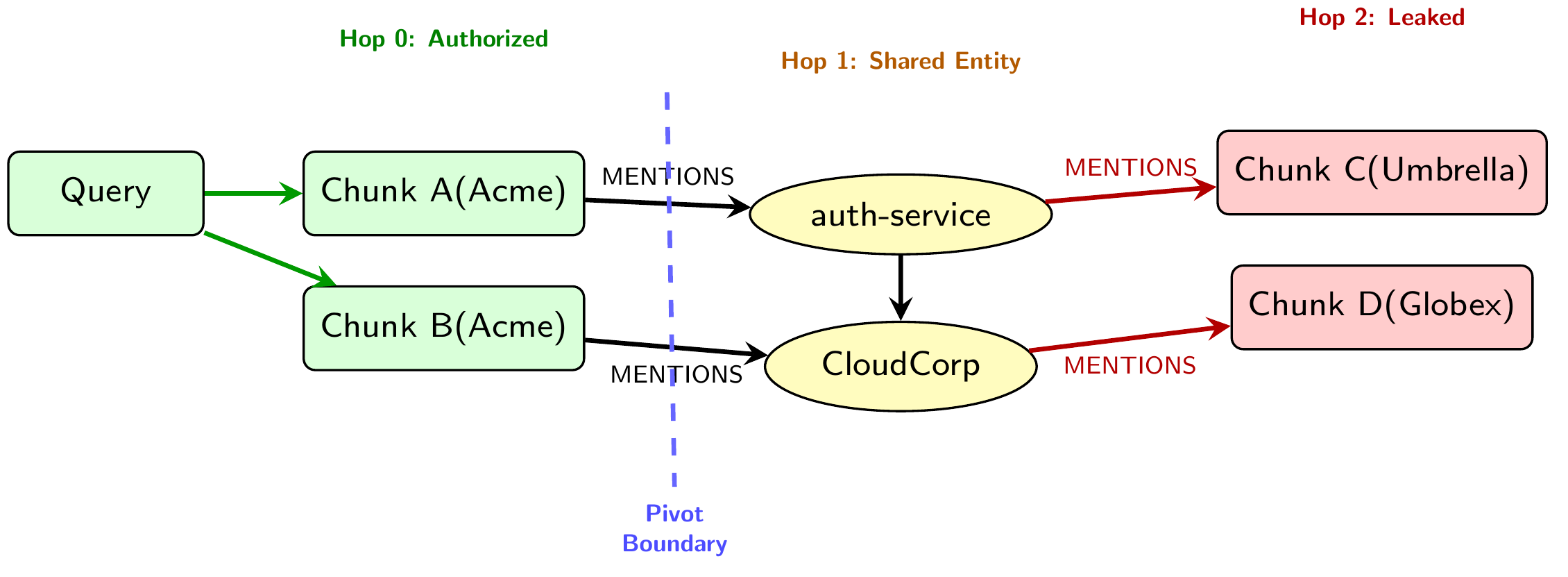}
\caption{The retrieval pivot attack surface. Vector retrieval returns
  authorized seed chunks (hop~0, green). BFS expansion traverses
  MENTIONS edges to shared entity nodes (hop~1, yellow), then onward
  to unauthorized chunks from other tenants (hop~2, red). The
  \textbf{pivot boundary} (dashed) is where authorization must be
  re-checked. D1 filters at this boundary, eliminating all leakage.}
\label{fig:pivot}
\end{figure}

\textbf{Running example.} Consider an INTERNAL-clearance engineer at
Acme Corp who queries ``What infrastructure does auth-service depend
on?'' The vector store returns authorized Acme engineering documents
mentioning auth-service. Entity linking maps ``auth-service'' to a
shared entity node in the knowledge graph. BFS expansion then walks:
auth-service $\to$ CloudCorp (shared vendor entity, hop 1) $\to$
Umbrella Security's CONFIDENTIAL incident-response documents (hop 2).
The engineer's authorized query silently pulls CONFIDENTIAL
cross-tenant data into the context---a 2-hop pivot through a
legitimate shared entity. Figure~\ref{fig:pivot} illustrates this
attack surface.

\section{Threat Model}
\label{sec:threat_model}

\subsection{System Model}

We consider a multi-tenant enterprise hybrid RAG system with the
following components:

\begin{itemize}
\item A \textbf{vector store} (ChromaDB) containing document chunk
  embeddings with tenant and sensitivity metadata.
\item A \textbf{knowledge graph} (Neo4j) containing entity nodes,
  chunk nodes, and typed edges (MENTIONS, DEPENDS\_ON, BELONGS\_TO,
  RELATED\_TO) extracted from the document corpus.
\item A \textbf{graph seeder} that maps vector-retrieved chunk IDs to
  their corresponding nodes in the knowledge graph.
\item A \textbf{graph expander} that performs BFS traversal from
  linked entities to depth $d$ (default $d=2$), implemented via
  Neo4j's \texttt{apoc.path.spanningTree} for hop-distance tracking.
\item Four organizational \textbf{tenants} with distinct data
  ownership boundaries.
\item Four \textbf{sensitivity tiers}: PUBLIC ($\approx$40\% of
  corpus), INTERNAL (30\%), CONFIDENTIAL (20\%), RESTRICTED (10\%).
\end{itemize}

\noindent\textbf{Authorization semantics.}
Chunk nodes inherit tenant and sensitivity labels from their source
documents at ingestion time; the vector store enforces these labels
during retrieval. Entity nodes, however, are \emph{label-free}:
because a single entity (e.g., \texttt{k8s-prod-cluster}) may be
mentioned by chunks from multiple tenants and sensitivity tiers,
production graph-construction pipelines~\cite{microsoftgraphrag2024,
hybridrag2024} do not assign ownership metadata to entity nodes.
This label gap is the structural root cause of the pivot
vulnerability: once graph expansion reaches an entity node, no
authorization check prevents traversal into chunks belonging to other
tenants or higher sensitivity tiers. Production knowledge graphs
typically rely on application-layer access control (RBAC/ABAC) rather
than graph-level enforcement, making this boundary particularly easy
to miss. Our defense D1
(\S\ref{sec:mitigations}) closes this gap by enforcing per-hop
authorization at every entity-to-chunk edge during expansion.

\subsection{Attacker Capabilities}

We consider two attacker models:

\textbf{Injection attacker.} The attacker holds a legitimate account
in one tenant (\texttt{acme\_engineering}) and can inject documents
into the shared corpus through standard channels---wiki edits, ticket
updates, shared document repositories. Injected documents undergo
normal ingestion: chunking, NER-based entity extraction, embedding,
and indexing into both vector and graph stores. The attacker's
injection budget is modest: 10--20 chunks, consistent with the budgets
shown effective in prior work~\cite{zou2025poisonedrag,
corruptrag2025}.

\textbf{No-injection attacker.} The attacker holds a legitimate
account and crafts queries that mention bridge entities (shared
infrastructure names, vendor references, cross-team personnel) to
maximize pivot probability through naturally occurring cross-tenant
entity connections. This attacker requires \emph{zero document
injection}---the organic structure of the multi-tenant knowledge graph
provides the pivot paths. Our evaluation shows that this attacker
model achieves $\text{RPR} = 0.95$ through benign queries alone
(\S\ref{sec:results}), demonstrating that the vulnerability is
\emph{structural} rather than injection-dependent.

\subsection{Attacker Goals}

\begin{itemize}
\item \textbf{G1: Cross-tenant access.} Cause the retrieval context
  for a query in the attacker's tenant to include chunks or entities
  belonging to other tenants.
\item \textbf{G2: Sensitivity escalation.} Cause an INTERNAL-clearance
  user's context to include CONFIDENTIAL or RESTRICTED items.
\item \textbf{G3: Amplified leakage.} Achieve leakage in the hybrid
  pipeline that exceeds what is possible with vector-only retrieval
  (i.e., $\text{AF} > 1$).
\end{itemize}

\subsection{Adaptive Attacker Considerations}

An adaptive attacker aware of D1 (per-hop authorization) might
attempt to bypass it through: (a) \textbf{metadata spoofing}---forging
tenant or sensitivity labels during document injection, or
(b) \textbf{same-tenant escalation}---crafting documents within the
attacker's own tenant that link to higher-sensitivity content within
the same tenant boundary. D1 is robust against (a) if and only if the
ingestion pipeline enforces metadata integrity (the metadata
is assigned by the system, not the uploader). We evaluate D1's
robustness under metadata corruption in \S\ref{sec:mislabel} and
discuss mitigation strategies in \S\ref{sec:metadata-integrity}.
Same-tenant escalation (b) targets \emph{sensitivity} boundaries
rather than \emph{tenant} boundaries and requires the attacker to
already have write access to the target tenant's corpus.

\subsection{Scope}

\textbf{We measure unauthorized items in the retrieval context
window}, not exfiltrated tokens or generated text. We focus exclusively
on the retrieval layer. We do not evaluate LLM generation quality,
jailbreaking, or prompt injection. The attacker does not have direct
database access, cannot modify model weights, and cannot alter the
pipeline configuration. We omit a graph-only baseline (P2) because
graph-only RAG without vector seeding is a fundamentally different
retrieval paradigm---our question is \emph{amplification created by
composition}, not graph-only security, which is addressed by prior
work~\cite{gragpoison2026, fewwords2025}.

\section{Retrieval Pivot Attacks}
\label{sec:attacks}

We define four non-adaptive attack strategies (A1--A4) that target
different stages of the hybrid pipeline---vector retrieval, entity
extraction, graph topology, and cross-tenant edges---and three
adaptive attacks (A5--A7) that target the defense mechanisms
themselves. Each strategy operates within the threat model of
\S\ref{sec:threat_model}: the attacker injects a small number of
crafted documents that undergo standard ingestion.

\subsection{A1: Seed Steering}

\textbf{Objective.} Maximize the probability that attacker-crafted
chunks are retrieved by target queries while embedding entity mentions
that link to sensitive graph neighborhoods.

\textbf{Mechanism.} The attacker estimates the query centroid for
target query families (e.g., ``infrastructure monitoring'') and crafts
10 chunks with high semantic overlap to this centroid. Each chunk
embeds 2 pivot entities---entities that are within 1--2 hops of
sensitive nodes in the knowledge graph (e.g., \texttt{k8s-prod-cluster},
\texttt{auth-service}). Chunks are labeled PUBLIC with low provenance
scores (0.3) to bypass any trust-based filtering. When retrieved,
entity linking maps the embedded entity mentions to graph nodes,
seeding expansion into sensitive neighborhoods.

\textbf{Entry point.} Vector retrieval (cosine similarity).\\
\textbf{Pivot mechanism.} Entity linking from retrieved chunk to
graph node.

\subsection{A2: Entity Anchor Injection}

\textbf{Objective.} Force entity linking to create dense connections
between injected chunks and sensitive graph neighborhoods.

\textbf{Mechanism.} The attacker identifies anchor entities adjacent
to sensitive nodes (e.g., entities 1 hop from restricted credentials)
and crafts chunks with dense entity mentions: 3+ mentions of the
primary target entity per chunk, plus 2 related entities. NER
extraction creates MENTIONS edges from each chunk to the target
entities. Any query that retrieves one of these chunks---even
incidentally---triggers graph expansion toward the sensitive area.

\textbf{Entry point.} Entity extraction during ingestion.\\
\textbf{Pivot mechanism.} MENTIONS edges from chunk to target entity
nodes.

\subsection{A3: Neighborhood Flooding}

\textbf{Objective.} Inflate the local density around a target entity
to increase the volume of sensitive content reachable through BFS
expansion.

\textbf{Mechanism.} The attacker injects 20 chunks, each mentioning
a high-value target entity near a sensitive subgraph. Each chunk also
mentions a different neighbor entity, creating a dense web of edges.
The result is a ``graph gravity well'': the target entity accumulates
many more edges than the graph average, causing BFS expansion to
gather a larger neighborhood when traversing through this entity.
Note that our pipeline uses uniform BFS traversal (not degree-weighted
or PageRank-based expansion), so the flooding increases reachable
volume rather than biasing traversal order.

\textbf{Entry point.} Graph topology manipulation via ingestion.\\
\textbf{Pivot mechanism.} Inflated local density increases BFS
expansion volume.

\subsection{A4: Bridge Node Attack}

\textbf{Objective.} Create artificial cross-tenant edges that enable
graph traversal from the attacker's tenant into a target tenant.

\textbf{Mechanism.} The attacker crafts 15 chunks that co-mention
entities from both the attacker's tenant and the target tenant within
the same document. NER extraction creates entity nodes on both sides
of the tenant boundary, and relation extraction creates RELATED\_TO
edges between them. After ingestion, BFS expansion from the
attacker's authorized subgraph can traverse these artificial bridge
edges into the target tenant's data.

\textbf{Entry point.} Cross-tenant entity co-mention.\\
\textbf{Pivot mechanism.} Artificial RELATED\_TO edges spanning tenant
boundaries.

\medskip

Algorithm~\ref{alg:pivot_attack} formalizes the general pivot attack
framework.

\begin{algorithm}[t]
\caption{General Retrieval Pivot Attack}
\label{alg:pivot_attack}
\begin{algorithmic}[1]
\Require Target query family $\mathcal{Q}_T$, injection budget $B$,
  pivot entities $E_p$, target neighborhood $N_T$
\Ensure Injected chunks $C_\text{atk}$
\For{$i = 1$ to $B$}
  \State $c_i \leftarrow$ \Call{CraftChunk}{$\mathcal{Q}_T$, $E_p$}
  \Comment{High similarity to $\mathcal{Q}_T$}
  \State $c_i.\text{entities} \leftarrow$ \Call{EmbedEntities}{$E_p$, $N_T$}
  \Comment{Mentions near $N_T$}
  \State $c_i.\text{sensitivity} \leftarrow$ PUBLIC
  \State $c_i.\text{provenance} \leftarrow 0.3$
  \State $C_\text{atk} \leftarrow C_\text{atk} \cup \{c_i\}$
\EndFor
\State \Call{Ingest}{$C_\text{atk}$}
\Comment{Standard pipeline: chunk, NER, embed, index}
\State \Return $C_\text{atk}$
\end{algorithmic}
\end{algorithm}

\section{Metrics}
\label{sec:metrics}

We introduce metrics to quantify retrieval pivot risk. Let $q$
denote a query, $u$ a user with tenant $t_u$ and clearance level
$\ell_u$, $S_k(q)$ the top-$k$ context set, and
$\text{Sensitive}(x, u)$ a predicate that is true when item $x$ has
sensitivity $> \ell_u$ or tenant $\neq t_u$ (entity nodes with
empty-string tenants are excluded from cross-tenant counts, as they
are tenant-neutral shared concepts---not leaked items themselves but
rather the pivot bridges through which leakage propagates). Let
$\text{Seeds}(q) \subseteq S_k(q)$ denote the set of entity nodes
that are linked from vector-retrieved chunks via NER extraction---these
are the graph entry points that initiate BFS expansion.

\textbf{Item serialization.} Each \emph{item} in $S_k(q)$ is a node
serialized for the LLM context: chunk nodes are serialized as their
source text; entity nodes are serialized as
\texttt{\{name, type, top-3 relations\}}. Context size counts the
total number of serialized nodes (chunks + entities) presented to
the LLM. The default pipeline configuration (depth $d=2$, branching
$\leq 10$, total $\leq 100$ nodes, top-$k{=}10$ vector results)
produces ${\sim}$110 items per query in the undefended hybrid
pipeline (P3). Table~\ref{tab:notation} summarizes the notation used
throughout.

\begin{table}[t]
\centering
\caption{Notation summary.}
\label{tab:notation}
\small
\begin{tabular}{l|l}
\toprule
\textbf{Symbol} & \textbf{Definition} \\
\midrule
$q$ & Query \\
$u$ & User with tenant $t_u$ and clearance $\ell_u$ \\
$S_k(q)$ & Top-$k$ context set (serialized items) \\
$\text{Seeds}(q)$ & Entity nodes linked from vector-retrieved chunks \\
$\text{Sensitive}(x,u)$ & $x$ has $\ell_x > \ell_u$ or $t_x \neq t_u$ \\
RPR & Prob.\ of any sensitive item in context \\
Leakage@k & Count of sensitive items in context \\
SWL & Severity-weighted leakage \\
AF($\epsilon$) & Leakage ratio: hybrid / max(vector, $\epsilon$) \\
PD & Min hops from seed to first sensitive node \\
\bottomrule
\end{tabular}
\end{table}

\subsection{Retrieval Pivot Risk (RPR)}

RPR measures the probability that a query's retrieval context contains
any unauthorized item:

\begin{equation}
\text{RPR}(u) = \Pr_{q \sim \mathcal{Q}}\left[\exists x \in S_k(q):
  \text{Sensitive}(x, u)\right]
\label{eq:rpr}
\end{equation}

\noindent RPR is operationalized as the fraction of evaluation queries
whose context contains at least one sensitive item. We report RPR
with 95\% bootstrap confidence intervals (10{,}000 resamples).

\subsection{Leakage@k}

Leakage@k counts the number of unauthorized items in the context:

\begin{equation}
\text{Leakage@k}(q, u) = \left|\{x \in S_k(q) :
  \text{Sensitive}(x, u)\}\right|
\label{eq:leakage}
\end{equation}

\noindent While RPR captures \emph{whether} leakage occurs,
Leakage@k captures its \emph{severity}. A context with 1 leaked item
and one with 20 both yield $\text{RPR} = 1$, but their Leakage@k
values differ by $20\times$.

We also define severity-weighted leakage (SWL) that weights violations by sensitivity gap (Appendix~\ref{app:swl}).

\subsection{Amplification Factor}

AF quantifies the leakage increase that hybrid retrieval introduces
compared to vector-only retrieval:
$\text{AF} = \mathbb{E}[\text{Leakage@k}]_{\text{hybrid}} /
\mathbb{E}[\text{Leakage@k}]_{\text{vector}}$.
When the vector baseline produces zero leakage (as our experiments
confirm), $\text{AF} = \infty$. We report a regularized variant
$\text{AF}(\epsilon) = \mathbb{E}[\text{Leakage@k}]_{\text{hybrid}} /
\max(\mathbb{E}[\text{Leakage@k}]_{\text{vector}}, \epsilon)$
with $\epsilon = 0.1$, alongside the absolute difference
$\Delta\text{Leakage}$. Because the vector baseline is zero,
$\Delta$Leakage is our primary magnitude measure; AF($\epsilon$)
serves as a secondary, plottable indicator.

\subsection{Pivot Depth (PD)}

PD measures the minimum graph distance from a seed node to the first
sensitive node reached during expansion:

\begin{equation}
\text{PD}(q) = \min\{d(s, x) : s \in \text{Seeds}(q),\;
  x \in S_k(q),\; \text{Sensitive}(x, u)\}
\label{eq:pd}
\end{equation}

\noindent where $d(s, x)$ is the shortest-path distance in the
knowledge graph from seed $s$ to node $x$, computed via
\texttt{apoc.path.spanningTree} which yields paths with explicit hop
counts. We report PD as a distribution (min, median, max) across
queries that exhibit leakage, rather than a single summary statistic.
Operationally, PD identifies the \emph{minimum traversal depth at
which enforcement must occur}: if all leakage has $\text{PD} = d$,
limiting expansion to depth $< d$ or inserting an authorization check
at depth $d$ is sufficient to prevent it.

\section{Experimental Setup}
\label{sec:setup}

\subsection{Synthetic Enterprise Corpus}

We generate a multi-tenant enterprise corpus of 1{,}000 documents
across four organizational tenants: \texttt{acme\_engineering},
\texttt{globex\_finance}, \texttt{initech\_hr}, and
\texttt{umbrella\_security}. Documents are produced by 12
domain-specific generators (3 per tenant) that embed realistic entity
mentions---system names, personnel, projects, compliance
standards---using curated entity pools.

Sensitivity tiers follow a realistic distribution: PUBLIC (40\%),
INTERNAL (30\%), CONFIDENTIAL (20\%), and RESTRICTED (10\%). Each
document includes ground-truth entity annotations and sensitivity
labels.

\textbf{Bridge entities.} We inject 15 bridge entities across 5
categories (shared vendors, shared infrastructure, shared personnel,
shared compliance standards, and shared projects) that naturally
appear in documents across multiple tenants. For example,
``CloudCorp'' appears in both engineering and finance documents,
and ``auth-service'' appears in both engineering and security
documents. After NER-based entity extraction, the knowledge graph
contains 40 naturally shared entities across tenant boundaries---not
through adversarial injection, but through legitimate cross-team
references that any multi-tenant organization would exhibit.

\subsection{Knowledge Graph Construction}

Documents undergo chunking (300-token windows, 50-token overlap),
producing ${\sim}$2{,}000 chunks from 1{,}000 source documents
(average ${\sim}$2 chunks per document). Chunks are processed through
spaCy NER extraction (\texttt{en\_core\_web\_sm}), two-pass
relation extraction (ground-truth relation resolution followed by
pattern-based extraction across 5 typed relations: DEPENDS\_ON,
OWNED\_BY, BELONGS\_TO, CONTAINS, DERIVED\_FROM, plus RELATED\_TO
fallback), and embedding via all-MiniLM-L6-v2 (384 dimensions). The
resulting knowledge graph contains 2{,}785 retrievable nodes (785
entity nodes + 2{,}000 chunk nodes; excluding 1{,}000 document
container nodes) and 15{,}514 edges (7{,}386 extracted relations
plus MENTIONS, CONTAINS, and BELONGS\_TO structural edges added during
graph construction).

\subsection{Enron Email Corpus}
\label{sec:enron}

To validate that Retrieval Pivot Risk is not an artifact of synthetic
construction, we evaluate on the Enron email corpus---a public-record
dataset of ${\sim}$500{,}000 corporate emails from ${\sim}$150 Enron
employees, released during the 2001 FERC investigation. We subsample
to 50{,}000 emails from the most active employees and assign tenants
based on departmental structure:

\begin{itemize}
\item \texttt{trading} --- Trading, West/East Power Trading
\item \texttt{legal} --- Legal, Government Affairs
\item \texttt{finance} --- Finance, Risk Management, Accounting
\item \texttt{energy\_services} --- Energy Services, Pipeline, ENA
\item \texttt{executive} --- Executive, Office of the Chairman
\end{itemize}

\noindent Sensitivity labels are assigned by keyword matching:
RESTRICTED (attorney-client privilege markers, password shares,
strategy memos), CONFIDENTIAL (deal negotiations, valuations, board
communications), INTERNAL (standard departmental communications),
PUBLIC (company-wide announcements). The resulting distribution
(92\% PUBLIC, 6.7\% CONFIDENTIAL, 0.7\% RESTRICTED, 0.3\% INTERNAL)
is skewed compared to the synthetic corpus---reflecting realistic
email traffic where most messages are routine.

Chunking (300-token windows, 50-token overlap) produces 152{,}064
chunks. spaCy NER extraction (\texttt{en\_core\_web\_sm}) identifies
2.07M entity mentions. The resulting knowledge graph contains
376{,}000 nodes (174{,}000 entity nodes + 152{,}000 chunk nodes +
50{,}000 document nodes) and 2.3M edges. Nineteen entities are
naturally shared across departmental boundaries---cross-department
executives (Ken Lay, Jeff Skilling, Andy Fastow), external
organizations (Arthur Andersen, Vinson \& Elkins), deal names
(Project Raptor, LJM2), and internal systems (EnronOnline).

\subsection{EDGAR 10-K Corpus}
\label{sec:edgar}

As a third corpus, we evaluate on SEC EDGAR 10-K annual reports---public
filings from 20 companies across four industry sectors (tech, finance,
healthcare, energy). Each sector serves as a tenant, and filing sections
receive sensitivity labels based on a Material Non-Public Information
(MNPI) framework: Items~1--4 (\textsc{public}), Items~7--8
(\textsc{internal}), Items~1A/7A/9A/11--12 (\textsc{confidential}),
Items~5/13 (\textsc{restricted}).

The EDGAR graph contains 19{,}527 nodes (887 documents, 3{,}692 chunks,
14{,}948 entities) and 427{,}415 edges. Cross-sector bridge entities
arise naturally from shared auditors (Big~4 firms), institutional
investors (BlackRock, Vanguard), and board members serving on companies
in multiple sectors. NER extraction on formal financial language produces
dense entity graphs ($\sim$4 entity mentions per chunk) but the
standardized vocabulary creates fewer distinctive cross-sector entity
connections than either the synthetic or Enron corpus.

\subsection{Pipeline Variants}

We evaluate 7 pipeline configurations (Table~\ref{tab:variants}):

\begin{table}[t]
\centering
\caption{Pipeline variants and their defense configurations.}
\label{tab:variants}
\small
\begin{tabular}{l|l|l}
\toprule
\textbf{ID} & \textbf{Pipeline} & \textbf{Defenses} \\
\midrule
P1 & Vector-only & Tenant prefilter \\
P3 & Hybrid baseline & None \\
P4 & Hybrid + D1 & Per-hop authz \\
P5 & Hybrid + D1,D2 & + Edge allowlist \\
P6 & Hybrid + D1--D3 & + Budgeted traversal \\
P7 & Hybrid + D1--D4 & + Trust weighting \\
P8 & Hybrid + D1--D5 & + Merge filter \\
\bottomrule
\end{tabular}
\end{table}

\noindent We omit a graph-only baseline (P2) because graph-only
retrieval without vector seeding is a fundamentally different
paradigm that does not involve the pivot boundary under study. Its
security properties are addressed by prior work on GraphRAG
attacks~\cite{gragpoison2026, fewwords2025}.

\subsection{Evaluation Protocol}

\textbf{Synthetic corpus.} We evaluate each pipeline on 500
template-generated queries: 350 benign queries (standard domain
questions stratified across 4 tenants and 3 clearance levels) and 150
adversarial queries (queries that mention bridge entities or target
cross-tenant pivot paths, stratified across attack types A1--A4). The
evaluation user belongs to \texttt{acme\_engineering} with INTERNAL
clearance.

\textbf{Enron corpus.} We evaluate P1, P3, and P4 on 200 queries
(100 benign + 100 adversarial) generated from department-specific
templates. The evaluation user belongs to \texttt{trading} with
INTERNAL clearance.

For each query, we measure RPR, Leakage@k, severity-weighted leakage,
AF($\epsilon$), $\Delta$Leakage, PD distribution, latency, and context
size. All RPR and Leakage@k values are reported with 95\% bootstrap
confidence intervals (10{,}000 resamples, seed 42). Graph expansion
uses \texttt{apoc.path.spanningTree} for BFS with hop-distance
tracking: each expanded node records its minimum distance from the
nearest seed, enabling precise PD measurement.

\textbf{Statistical methodology.}
All metrics are reported with 95\% bootstrap confidence intervals
(10{,}000 resamples, percentile method, seed 42). We use non-parametric
bootstrap because RPR values near 0 or 1 violate normal approximations,
and leakage distributions are highly skewed. Our query sets (500
synthetic, 200 Enron) each provide power $> 0.80$ for detecting 5pp
differences (Appendix~\ref{sec:stats-appendix} provides sample size
justification, multiple comparison corrections, and $\epsilon$
sensitivity analysis).

\textbf{Reproducibility.} All experiments use a single fixed random
seed (42) for corpus generation, query sampling, and bootstrap
resampling.  Pipeline configurations are built programmatically (not
from YAML) to ensure parameter consistency across variants.
Appendix~\ref{sec:artifact} provides the full reproduction sequence;
Appendix~\ref{sec:corpus-details} details the corpus generator;
Appendix~\ref{sec:query-templates} lists all query templates.
The codebase includes 255 passing unit tests.

\subsection{Attack Evaluation}

We evaluate all four attacks (A1--A4) against the undefended hybrid
pipeline (P3) and three defense configurations (P4, P6, P8). For each
attack, payloads are injected into a clean corpus, and 10 adversarial
queries are executed against each pipeline variant. After each attack
evaluation, the graph is rebuilt from the clean corpus to prevent
cross-contamination between attack experiments.

\section{Results}
\label{sec:results}

\subsection{Hybrid RAG Amplifies Leakage}

Table~\ref{tab:defense-ablation} presents security metrics across all
pipeline variants. The central finding is clear: \textbf{the
undefended hybrid pipeline (P3) exhibits high leakage rates while the
vector-only baseline (P1) leaks nothing.}

P1 achieves $\text{RPR} = 0.0$ on both query sets---the vector
store's tenant prefilter prevents any cross-tenant or
sensitivity-escalated content from reaching the context. In contrast,
P3 achieves $\text{RPR} = 0.954$ [95\% CI: 0.931, 0.974] on 350
benign queries (334 of 350 queries produce leakage) and
$\text{RPR} = 0.947$ [0.907, 0.980] on 150 adversarial queries (142
of 150 queries leak). Mean Leakage@k reaches 16.0 items for benign
queries and 19.4 for adversarial queries---meaning that roughly
15--18\% of the 110-item context consists of unauthorized content.
Severity-weighted leakage averages 22.9 (benign) and 26.4
(adversarial), indicating that leaked items skew toward higher
sensitivity tiers.

\begin{table*}[t]
\centering
\caption{Defense ablation: security metrics across pipeline variants
  (500 queries: 350 benign + 150 adversarial). RPR = Retrieval Pivot
  Risk with 95\% bootstrap CI, Leak = mean Leakage@k, SWL =
  severity-weighted leakage, PD = Pivot Depth (hops), Ctx = mean
  context size. ``--'' indicates no leakage occurred (PD undefined).}
\label{tab:defense-ablation}
\small
\begin{tabular}{l|ccccc|ccccc}
\toprule
 & \multicolumn{5}{c|}{\textbf{Benign Queries ($n=350$)}} &
   \multicolumn{5}{c}{\textbf{Adversarial Queries ($n=150$)}} \\
\textbf{Variant} & RPR [CI] & Leak & SWL & PD & Ctx &
  RPR [CI] & Leak & SWL & PD & Ctx \\
\midrule
P1 (Vector) & 0.000 & 0.0 & 0.0 & -- & 10 &
  0.000 & 0.0 & 0.0 & -- & 10 \\
P3 (Hybrid) & .954 [.931,.974] & 16.0 & 22.9 & 2.0 & 110 &
  .947 [.907,.980] & 19.4 & 26.4 & 2.0 & 110 \\
P4 (+D1) & 0.000 & 0.0 & 0.0 & -- & 56 &
  0.000 & 0.0 & 0.0 & -- & 50 \\
P5 (+D1,D2) & 0.000 & 0.0 & 0.0 & -- & 57 &
  0.000 & 0.0 & 0.0 & -- & 51 \\
P6 (+D1--D3) & 0.000 & 0.0 & 0.0 & -- & 29 &
  0.000 & 0.0 & 0.0 & -- & 28 \\
P7 (+D1--D4) & 0.000 & 0.0 & 0.0 & -- & 28 &
  0.000 & 0.0 & 0.0 & -- & 24 \\
P8 (All) & 0.000 & 0.0 & 0.0 & -- & 20 &
  0.000 & 0.0 & 0.0 & -- & 20 \\
\bottomrule
\end{tabular}
\end{table*}

Because the vector baseline (P1) produces zero leakage, the absolute
difference $\Delta\text{Leakage} = 16.0$ (benign) and $19.4$
(adversarial) is the most direct measure of the hybrid penalty:
each query exposes 16--19 additional unauthorized items purely
from graph expansion.  The classical Amplification Factor is formally
$\text{AF} = \infty$ (division by zero); the regularized variant
$\text{AF}(\epsilon{=}0.1) = 160$--$194\times$ provides a finite
proxy.\footnote{$\text{AF}(\epsilon)$ is sensitive to the choice of
$\epsilon$; see Appendix~\ref{sec:stats-appendix} for a robustness
check across $\epsilon \in \{0.01, 0.05, 0.1, 0.5\}$.}
The hybrid architecture does not merely amplify existing risk---it
creates risk from nothing: P1 achieves $\text{RPR} = 0.0$ while P3
reaches $\text{RPR} = 0.95$ on identical queries.

\begin{figure}[t]
\centering
\includegraphics[width=\columnwidth]{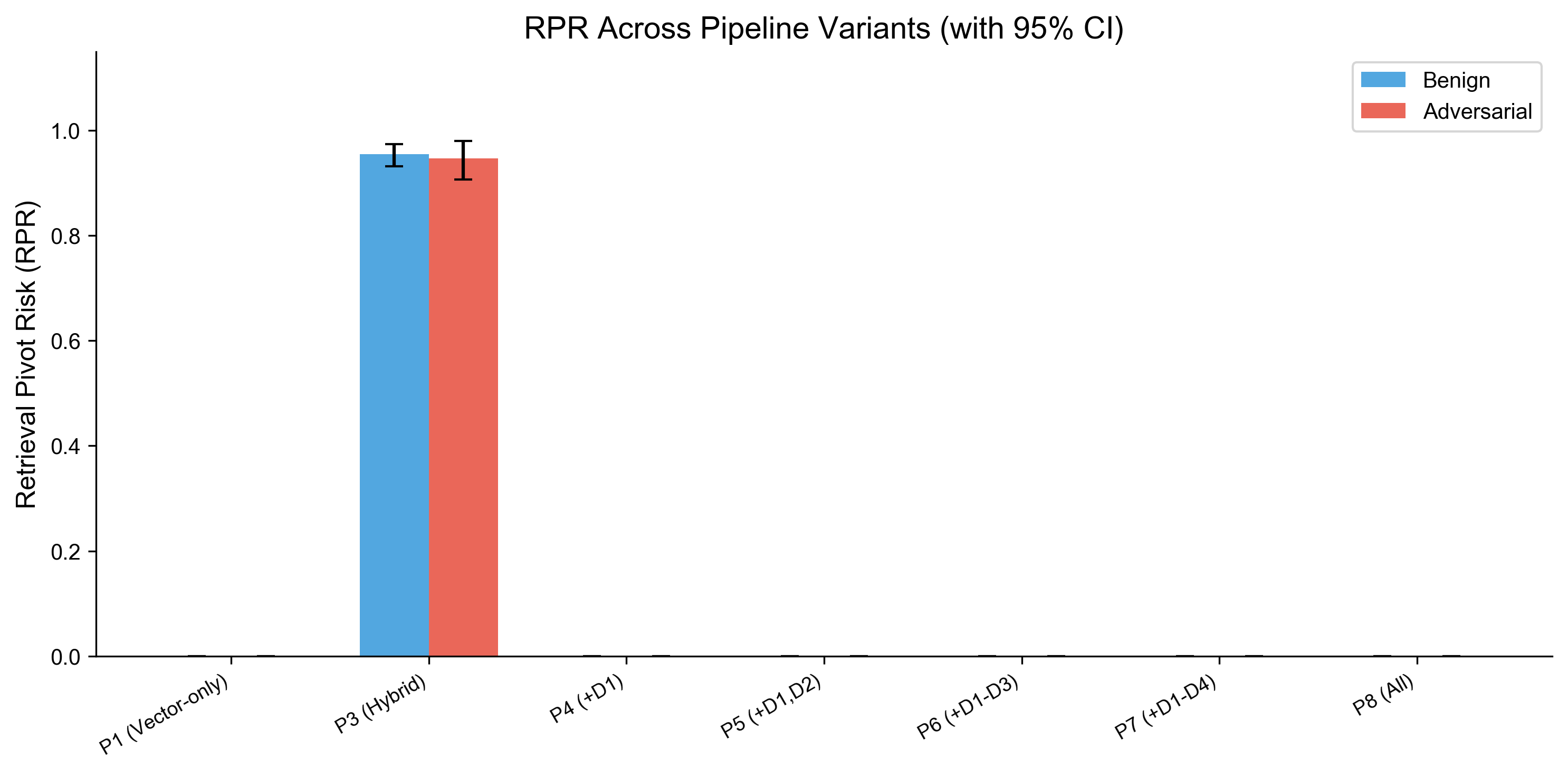}
\caption{Retrieval Pivot Risk with 95\% bootstrap CIs across pipeline
  variants. P3 (undefended hybrid) shows
  $\text{RPR} \approx 0.95$. All defended variants (P4--P8) achieve
  $\text{RPR} = 0.0$.}
\label{fig:rpr}
\end{figure}

Figure~\ref{fig:rpr} shows RPR with bootstrap confidence intervals
across pipeline variants.

\subsection{The PD = 2 Structural Signature}

All leakage in P3 occurs at exactly $\text{PD} = 2$ hops: the PD
distribution is $(\min=2, \text{median}=2, \max=2)$ across all 476
leaking queries. This uniformity is not coincidental---it is a
structural consequence of the bipartite chunk-entity graph topology in
our construction. The pivot path is:

\begin{enumerate}
\item \textbf{Hop 0}: Authorized seed chunk (retrieved by vector
  similarity, passes tenant prefilter).
\item \textbf{Hop 1}: Shared entity node (linked via NER from seed
  chunk text; entity nodes carry no tenant ownership).
\item \textbf{Hop 2}: Unauthorized chunk (connected to the shared
  entity via MENTIONS edge; belongs to a different tenant or higher
  sensitivity tier).
\end{enumerate}

This 2-hop pattern is inherent to any hybrid RAG system that
constructs a bipartite graph between chunks and entities with
expansion depth $d \geq 2$: the entity-linking step creates a
structural bridge between vector-retrieved content and graph-stored
content. The PD=2 finding is \emph{structural in our bipartite
construction}---knowledge graphs with richer entity-to-entity
relationships (e.g., ontological hierarchies, multi-hop inference
chains) may exhibit leakage at deeper pivot depths. We discuss
how PD varies with traversal parameters in
\S\ref{sec:traversal-sweep}.

\subsection{Organic Leakage: No Injection Required}
\label{sec:organic}

A critical observation from our 350 benign queries is that
$\text{RPR} = 0.954$ \emph{without any adversarial injection}. The
40 naturally shared entities across tenant boundaries---shared
infrastructure, vendors, personnel, compliance standards, and
projects---provide sufficient pivot paths for massive leakage through
ordinary queries. In our corpus, 334 of 350 benign queries (95.4\%)
trigger cross-tenant leakage through organic entity connections. This
means the vulnerability is \emph{structural}: it exists in any
multi-tenant hybrid RAG deployment where tenants share real-world
entities, regardless of whether an attacker injects content.

\textbf{Bridge category analysis.} To understand which types of
shared entities drive leakage, we analyze the hop-1 pivot nodes in
each leaking query's traversal path and classify them against the 5
bridge entity categories (Table~\ref{tab:organic-leakage}).

\begin{table}[t]
\centering
\caption{Organic leakage by bridge entity category under P3 (no
  injection). Queries = leaking queries with that bridge type at
  hop 1. Leak = attributed leaked items.}
\label{tab:organic-leakage}
\small
\begin{tabular}{l|rr|rr}
\toprule
\textbf{Bridge Category} & \multicolumn{2}{c|}{\textbf{Benign}} &
  \multicolumn{2}{c}{\textbf{Adversarial}} \\
 & Queries & Leak & Queries & Leak \\
\midrule
  Personnel & 104 & 1{,}321 & 67 & 1{,}244 \\
  Compliance & 46 & 782 & 33 & 489 \\
  Infrastructure & 43 & 406 & 16 & 213 \\
  Vendor & 0 & 0 & 30 & 218 \\
  Project & 7 & 87 & 24 & 228 \\
\midrule
  (No bridge at hop 1) & 175 & --- & 22 & --- \\
\midrule
  Total leaking & 334 & 5{,}595 & 142 & 2{,}907 \\
\bottomrule
\end{tabular}
\end{table}

Personnel entities (shared employees like ``Maria Chen'') dominate:
they appear at hop 1 in 31\% of benign leaking queries and 47\% of
adversarial leaking queries, accounting for 23.6\% and 42.8\% of
attributed leakage respectively. Compliance entities (SOC2, PCI-DSS,
ISO27001) rank second. Notably, 52\% of benign leaking queries have
\emph{no recognized bridge entity} at hop 1---non-bridge entities
(monetary amounts, dates, generic organizational terms extracted by
spaCy NER) also create unintended cross-tenant paths, indicating
that the leakage risk is broader than just named shared entities.

\paragraph{Connectivity sensitivity.}
\label{sec:connectivity}
RPR remains stable (0.93--0.95) regardless of bridge entity count; mean leakage scales monotonically from 21.5 (0 bridges) to 34.5 (40 bridges), confirming that bridge entities amplify volume but do not enable leakage (Appendix~\ref{app:connectivity}).

\paragraph{Embedding model sensitivity.}
\label{sec:embedding}
Replacing all-MiniLM-L6-v2 (384d) with all-mpnet-base-v2 (768d) actually \emph{increases} RPR (0.954$\to$0.994 benign, 0.947$\to$0.980 adversarial), confirming the vulnerability is structural and model-independent (Appendix~\ref{app:embedding}).

\subsection{Attack Evaluation}
\label{sec:attack-eval}

Table~\ref{tab:attack-heatmap} presents RPR under each attack across
pipeline variants. All four attacks achieve $\text{RPR} = 1.0$ against
the undefended hybrid pipeline (P3)---every adversarial query produces
cross-tenant leakage. Critically, all four attacks achieve
$\text{RPR} = 0.0$ against every defended variant (P4, P6, P8).

\begin{table}[t]
\centering
\caption{RPR under each attack type across pipeline variants.
  Mean Leakage@k shown in parentheses. All attacks fail against
  D1-defended pipelines.}
\label{tab:attack-heatmap}
\small
\begin{tabular}{l|cccc}
\toprule
\textbf{Attack} & \textbf{P3} & \textbf{P4} & \textbf{P6} &
  \textbf{P8} \\
\midrule
  A1 (Seed Steer) & 1.00 (20.5) & 0.00 (0.0) & 0.00 (0.0) &
    0.00 (0.0) \\
  A2 (Entity Anchor) & 1.00 (20.5) & 0.00 (0.0) & 0.00 (0.0) &
    0.00 (0.0) \\
  A3 (Nbhd Flood) & 1.00 (20.5) & 0.00 (0.0) & 0.00 (0.0) &
    0.00 (0.0) \\
  A4 (Bridge Node) & 1.00 (20.5) & 0.00 (0.0) & 0.00 (0.0) &
    0.00 (0.0) \\
\bottomrule
\end{tabular}
\end{table}

Table~\ref{tab:attack-details} provides injection details for each
attack. The attacks span a range of injection budgets (9--20 chunks)
and entity strategies (1--2 target entities, 1--6 MENTIONS edges
created).

\begin{table}[t]
\centering
\caption{Attack injection details. All attacks achieve RPR=1.0 on P3
  and RPR=0.0 on P4/P6/P8.}
\label{tab:attack-details}
\small
\begin{tabular}{l|cccc}
\toprule
\textbf{Attack} & \textbf{Payloads} & \textbf{Chunks} &
  \textbf{Entities} & \textbf{Mentions} \\
\midrule
  A1 & 9 & 9 & 1 & 3 \\
  A2 & 10 & 10 & 1 & 6 \\
  A3 & 20 & 20 & 1 & 4 \\
  A4 & 15 & 15 & 2 & 6 \\
\bottomrule
\end{tabular}
\end{table}

Two aspects of this uniformity are themselves findings.
\textbf{Leakage is bounded by the expansion window, not the attack
mechanism}: all four attacks produce identical Leakage@k
($\approx$20.5) against P3 because the bottleneck is the
\emph{unguarded graph traversal}---all four simply steer queries into
the same undefended 2-hop expansion path, and the total\_nodes budget
(100) caps the leaked volume. This means more sophisticated injection
strategies yield no additional leakage beyond what organic entity
overlap already provides. Second, D1 blocks \emph{all} four strategies
because it operates on node \emph{properties} (tenant, sensitivity)
rather than graph \emph{structure} (paths, edges). The attacks
manipulate paths---optimizing similarity (A1), creating dense
connections (A2), inflating density (A3), or bridging boundaries
(A4)---but D1 filters on properties at each hop, making the path
irrelevant.

\paragraph{Adaptive attacks (A5--A7).}
\label{sec:adaptive-eval}
Three adaptive attacks---metadata forgery (A5, up to 10\%), entity manipulation (A6), and query manipulation (A7)---all achieve $\text{RPR} = 1.0$ against P3 but $\text{RPR} = 0.0$ against all defended pipelines, because D1 filters on node properties rather than graph structure (Appendix~\ref{app:adaptive}).

\paragraph{Generation impact.}
\label{sec:generation-impact}
Evaluating three production LLMs across all three corpora confirms that retrieval-level leakage translates to generation-level contamination: Entity Contamination Rates reach 0.32 (Claude Sonnet 4.5, synthetic) and decrease with entity distinctiveness across corpora (Appendix~\ref{app:generation}).

\paragraph{Traversal parameter sweep.}
\label{sec:traversal-sweep}
A 27-configuration sweep (depth $\times$ branching $\times$ total node budget) shows that total node budget is the primary leakage control ($\text{total\_nodes} \leq 25$ eliminates leakage), depth must be $\geq 2$ for the pivot to occur, and branching factor is irrelevant given a total node cap (Appendix~\ref{app:traversal}).

\paragraph{Latency and overhead.}
D1 adds $<$1ms latency overhead (P3 p50 26.7ms $\to$ P4 26.5ms); the full defense stack (P8) is actually \emph{faster} than undefended P3 (22.7 vs.\ 26.7ms p50) because budgeted traversal reduces expansion volume (Appendix~\ref{app:latency}).

\paragraph{Metadata integrity stress test.}
\label{sec:mislabel}
D1 maintains $\text{RPR} = 0.0$ under random sensitivity mislabeling up to 5\%, because its primary protection is the tenant filter, which is unaffected by sensitivity corruption (Appendix~\ref{app:mislabel}).

\subsection{Defense Ablation}
\label{sec:defense-ablation}

\begin{figure}[t]
\centering
\includegraphics[width=\columnwidth]{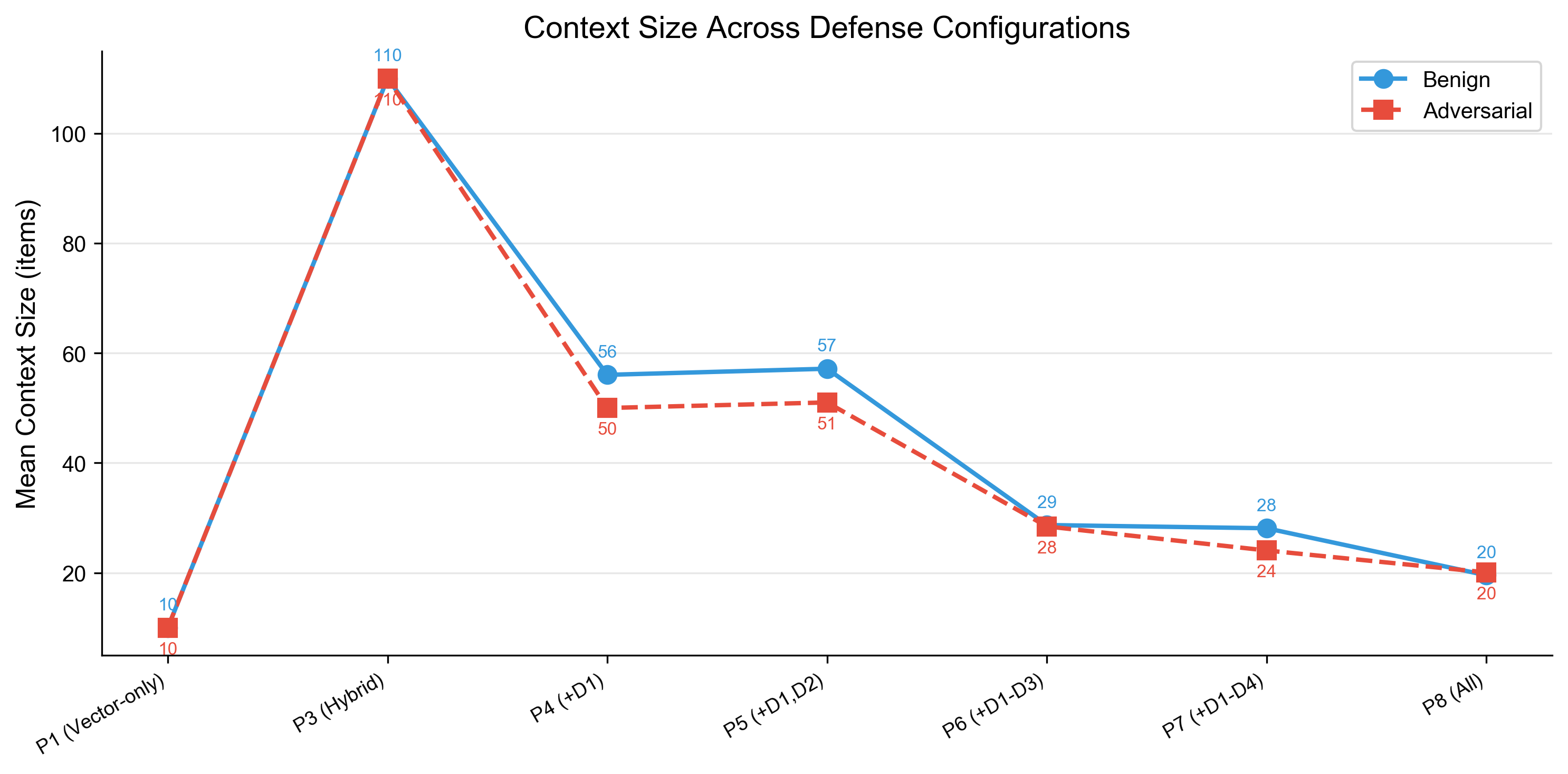}
\caption{Mean context size under progressive defenses. D1 alone
  reduces context from 110 to 50--56 items (removing unauthorized
  content). D3--D5 further reduce noise, reaching 19--20 items with
  the full defense stack.}
\label{fig:context}
\end{figure}

The defense stack shows a clear pattern
(Figure~\ref{fig:context}):

\begin{itemize}
\item \textbf{D1 alone} (P4): RPR drops from $0.95$ to $0.0$.
  Context reduces from 110 to 50--56 items (the unauthorized items
  are removed; authorized graph-expanded content is retained).
\item \textbf{D1 + D2} (P5): No additional security improvement;
  context remains similar (51--57). Edge allowlisting provides defense
  in depth but does not further reduce leakage already eliminated
  by D1.
\item \textbf{D1--D3} (P6): Context drops to 28--29. Budgeted
  traversal caps the number of expanded nodes, reducing context noise
  from authorized but irrelevant graph content.
\item \textbf{D1--D4} (P7): Context drops to 24--28.
  Trust-weighted filtering removes low-provenance nodes.
\item \textbf{D1--D5} (P8): Context reaches 19--20 items. The merge
  filter provides a final defense-in-depth check, reducing context
  by 82\% from the undefended baseline.
\end{itemize}

The key insight is that D1 is both \emph{necessary and sufficient}
for security (eliminating all leakage), while D2--D5 serve as
\emph{utility optimizers} (reducing context noise and providing
defense in depth).

\paragraph{Utility impact.}
\label{sec:utility}
D1 (P4) retains 56 authorized items per query---5.6$\times$ more than vector-only (P1)---while eliminating all leakage. The full stack (P8) retains 20 items (2$\times$ vector-only). D3--D5 serve as context quality optimizers rather than security controls (Appendix~\ref{app:utility}).

\subsection{Cross-Dataset Validation}
\label{sec:cross-dataset}

To verify that Retrieval Pivot Risk is not an artifact of synthetic
corpus construction, we repeat the baseline evaluation (P1, P3, P4)
on the Enron email corpus (\S\ref{sec:enron}) and EDGAR 10-K filings
(\S\ref{sec:edgar}). Table~\ref{tab:cross-dataset} compares results
across all three corpora.

\begin{table}[t]
\centering
\caption{Cross-dataset baseline: retrieval security metrics for undefended (P3)
and defended (P4) hybrid RAG across three corpora.
RPR = Retrieval Pivot Risk, Leak = mean leaked items per query,
PD = Pivot Depth (hops), Ctx = mean context size.}
\label{tab:cross-dataset}
\small
\begin{tabular}{ll|cccc}
\toprule
\textbf{Dataset} & \textbf{Pipeline} & \textbf{RPR} & \textbf{Leak} & \textbf{PD} & \textbf{Ctx} \\
\midrule
  \multirow{3}{*}{Synthetic} & P1 (Vector) & 0.000 & 0.0 & -- & 10 \\
                             & P3 (Hybrid) & 0.954 & 16.0 & 2.0 & 110 \\
                             & P4 (+D1)    & 0.000 & 0.0 & -- & 56 \\
\midrule
  \multirow{3}{*}{Enron}     & P1 (Vector) & 0.000 & 0.0 & -- & 10 \\
                             & P3 (Hybrid) & 0.695 & 7.1 & 2.0 & 45 \\
                             & P4 (+D1)    & 0.000 & 0.0 & -- & 25 \\
\midrule
  \multirow{3}{*}{EDGAR}     & P1 (Vector) & 0.000 & 0.0 & -- & 10 \\
                             & P3 (Hybrid) & 0.085 & 0.4 & 2.0 & 39 \\
                             & P4 (+D1)    & 0.000 & 0.0 & -- & 22 \\
\bottomrule
\end{tabular}
\vspace{2pt}
\begin{minipage}{0.95\linewidth}
\footnotesize
\textit{Note:} Synthetic: 1{,}000 documents, 4 tenants, 40 bridge entities.
Enron: 50{,}000 emails, 5 departments, 19 cross-department entities.
EDGAR: 887 10-K sections, 4 sector-tenants, 13 cross-sector entities.
\end{minipage}
\end{table}

Four findings emerge from the cross-dataset comparison:

\textbf{The vulnerability generalizes.} Enron's undefended hybrid
pipeline (P3) produces $\text{RPR} = 0.695$ on benign queries---139
of 200 queries leak cross-department content, with a mean of 7.1
leaked items per query. EDGAR produces $\text{RPR} = 0.085$---lower
but still affecting 17 of 200 queries. The vulnerability is not an
artifact of synthetic construction: real corporate data with natural
organizational boundaries exhibits the same pivot mechanism.

\textbf{RPR scales with entity connectivity.} The three corpora span
a wide range: synthetic ($\text{RPR} = 0.954$, 40 bridge entities),
Enron ($\text{RPR} = 0.695$, 19 shared entities), and EDGAR
($\text{RPR} = 0.085$, 13 cross-sector entities). RPR correlates
with the number and density of cross-tenant entity connections. The
synthetic graph has curated cross-references, Enron has informal
cross-department email traffic, and EDGAR's formal financial language
creates fewer organic entity connections across sectors. This confirms
that RPR is a function of entity connectivity density, not corpus size.

\textbf{D1 eliminates all leakage on all three corpora.} P4 achieves
$\text{RPR} = 0.0$ on Enron (context size 25), EDGAR (context size
22), and synthetic (context size 56). The defense generalizes without
modification: the same per-hop tenant-and-sensitivity check eliminates
leakage regardless of corpus size, entity distribution, or graph
topology.

\textbf{PD = 2 persists.} All leakage in all three corpora occurs at
exactly $\text{PD} = 2$, confirming that the bipartite pivot
(chunk $\to$ entity $\to$ chunk) is a structural invariant of hybrid
RAG systems that construct knowledge graphs via entity linking,
independent of the underlying document domain.

\section{Mitigations}
\label{sec:mitigations}

We propose five defenses that operate at different stages of the
hybrid pipeline, forming a defense-in-depth architecture
(Table~\ref{tab:defenses}).

\begin{table}[t]
\centering
\caption{Defense mechanisms, pipeline integration points, and
  experimental impact.}
\label{tab:defenses}
\small
\begin{tabular}{l|l|l}
\toprule
\textbf{Defense} & \textbf{Stage} & \textbf{Effect} \\
\midrule
D1: Per-hop authz & Post-expansion & RPR $\to$ 0.0 \\
D2: Edge allowlist & Traversal query & Defense in depth \\
D3: Budget & Traversal params & Ctx 110 $\to$ 28 \\
D4: Trust weight & Post-expansion & Low-prov removal \\
D5: Merge filter & Post-merge & Final backstop \\
\bottomrule
\end{tabular}
\end{table}

\subsection{D1: Per-Hop Authorization}

Existing hybrid RAG frameworks (LangChain, LlamaIndex, Microsoft
GraphRAG) enforce authorization only during vector retrieval, implicitly
assuming that downstream graph expansion operates on the same filtered
dataset. This assumption fails because entity linking crosses the
authorization boundary: the graph traversal engine has access to the
\emph{entire} knowledge graph, not only the tenant-filtered subset.

Per-hop authorization is an \textbf{intentionally conservative
minimum-viable guardrail}: it re-checks access control predicates on
every node reached during graph expansion. For each expanded node $v$, the
defense evaluates:

\begin{equation}
\text{auth}(u, v) = (v.\text{tenant} = t_u) \wedge
  (v.\text{sensitivity} \leq \ell_u)
\end{equation}

Nodes failing this check are removed from the expansion result before
context assembly. In our implementation, D1 operates as a
\textbf{post-expansion filter}: the BFS traversal runs
unrestricted, and the authorization check is applied to the result
set. This is simpler to implement than in-traversal pruning
(which would stop expansion at unauthorized nodes) but is less
efficient because it expands nodes that will be discarded. A true
per-hop pruning implementation would stop expansion at unauthorized
nodes, preventing their children from being explored. We chose
post-expansion filtering for implementation simplicity; the security
properties are identical (both produce $\text{RPR} = 0.0$), but
in-traversal pruning would further reduce latency and context
processing overhead.

D1 is the most effective defense: it reduces RPR from 0.95 to 0.0
across all query types and all attack variants, with $<$1ms latency
overhead.

\paragraph{Entity tenant semantics.}
A subtle but critical design decision concerns entity nodes. In our
knowledge graph, entity nodes (extracted via NER) represent shared
concepts---people, systems, vendors---that may appear in documents
across multiple tenants. These nodes carry \texttt{tenant = ""}
(empty string) rather than any specific tenant label, because they
are \emph{tenant-neutral}: the concept ``CloudCorp'' belongs to no
single tenant. Under D1's authorization check, empty-tenant nodes
fail the tenant-match predicate ($v.\text{tenant} \neq t_u$) and are
filtered from the expansion result. This means D1 removes
\emph{all entity nodes} from the context---not just unauthorized ones.

This entity-level filtering is precisely the mechanism by which D1
eliminates cross-tenant leakage. The 2-hop pivot path (chunk $\to$
entity $\to$ chunk) requires traversal through a shared entity node.
By filtering entity nodes, D1 severs this path at hop 1, preventing
the traversal from reaching the unauthorized chunk at hop 2. The
trade-off is that entity information (which may be useful for answer
generation) is excluded from the final context. We quantify this
impact in Section~\ref{sec:discussion}.

We acknowledge that D1's entity filtering is functionally equivalent
to disabling entity traversal---a valid criticism. However, this is
the \emph{minimum viable defense}, not the optimal one. A
finer-grained \textbf{entity-aware authorization} scheme would: (a)
label each entity with the set of tenants whose documents mention it
(e.g., ``CloudCorp'' $\to$ \{acme, globex\}), (b) allow traversal
\emph{through} entities whose tenant set includes $t_u$, and (c) apply
the authorization check only at \emph{chunk} nodes (the terminal
nodes that carry sensitive text). This design would preserve entity
context for within-tenant traversals while still blocking cross-tenant
chunk access. We leave its implementation and security analysis
to future work, and position D1 as the conservative baseline that
demonstrates the \emph{existence} of the boundary problem.

The critical insight is that \emph{existing graph databases already
store the metadata needed for this check} (tenant labels, sensitivity
tiers). The defense does not require new infrastructure---only the
discipline to re-check authorization at the graph layer rather than
relying solely on the vector prefilter.

\subsection{D1 Metadata Integrity Assumption}
\label{sec:metadata-integrity}

D1 assumes that node metadata (tenant, sensitivity) is trustworthy.
If an attacker can forge metadata during document injection, D1 can
be bypassed. Our metadata stress test (\S\ref{sec:mislabel}) shows
that D1 is robust to random mislabeling up to 5\%, but targeted
metadata forgery is a different threat. Mitigations include:
(a) system-assigned metadata during ingestion (the uploader cannot
choose tenant or sensitivity labels), (b) cryptographic signing of
metadata at ingestion time, and (c) periodic metadata audits that
compare node labels against source-of-truth identity systems. In
practice, most enterprise document management systems already enforce
system-assigned tenant labels, making (a) the natural deployment
model. Our stress test confirms D1 maintains $\text{RPR} = 0.0$ at mislabel rates up to 5\% (Appendix~\ref{app:mislabel}). Targeted metadata forgery could bypass this defense if ingestion pipelines allow user-controlled labels; deployments should therefore treat metadata assignment as a system-only operation.

\subsection{D2: Edge Allowlist}

Edge allowlisting restricts graph traversal to pre-approved edge
types per query class. The expander's BFS query includes a
\texttt{relationshipFilter} parameter that limits which edge types
can be traversed. For example, general queries may traverse MENTIONS,
DEPENDS\_ON, and BELONGS\_TO edges, while RELATED\_TO edges (which
are the primary vector for bridge attacks) are excluded.

\subsection{D3: Budgeted Traversal}

Budgeted traversal enforces hard limits on BFS expansion: maximum hop
depth, maximum branching factor per node, and maximum total expanded
nodes. Our traversal sweep (\S\ref{sec:traversal-sweep}) shows that
$\text{total\_nodes} \leq 25$ eliminates leakage even without D1,
while the default budget ($d_\text{max}=2$, branching $\leq 10$,
total $\leq 50$) reduces context from 56 items (D1 only) to 28 items,
removing authorized but irrelevant graph content that adds noise.

\subsection{D4: Trust-Weighted Expansion}

Trust-weighted expansion filters expanded nodes by their provenance
score. Each node carries a provenance score ($0.0$--$1.0$) based on
its source reliability. The defense applies a minimum threshold
(default: 0.6), removing low-provenance content. This is particularly
effective against injected attack payloads, which carry provenance
scores of 0.3--0.4 in our attack implementations.

\subsection{D5: Merge-Time Policy Filter}

The merge-time policy filter performs a final access control check
after vector and graph results are combined. This defense acts as a
backstop: even if earlier defenses miss an unauthorized item, the
merge filter removes any item whose sensitivity exceeds the user's
clearance before the context reaches the LLM.

\section{Discussion}
\label{sec:discussion}

\subsection{Security-Utility Tradeoff}

Our results reveal a favorable security-utility tradeoff. D1 alone
achieves complete security ($\text{RPR} = 0.0$) while retaining 50\%
of the graph-expanded context ($110 \to 50\text{--}56$ items). The
retained items are authorized content that contributes to answer
quality. Adding D3--D5 further reduces context to 19--20
items---an 82\% reduction from the undefended baseline---but the
additional defenses trade context volume for noise reduction rather
than security improvement.

D1's entity over-filtering trade-off is analyzed in detail in \S\ref{sec:mitigations} (entity tenant semantics).

The practical recommendation is clear: \textbf{deploy D1 immediately
as a minimum viable defense.} Add D3 and D4 if context noise is
impacting LLM generation quality. D2 and D5 provide defense in
depth for compliance-sensitive deployments. Practically, this means
systems should treat tenant-neutral entity nodes as \emph{privileged
pivot infrastructure} that must carry explicit authorization semantics
(or be excluded) in multi-tenant deployments; leaving them unlabeled
creates a structurally inevitable cross-tenant pivot path.

\subsection{Why PD = 2 Across All Three Corpora}

The uniform $\text{PD} = 2$ finding across all three corpora (2.8K--376K nodes) follows directly from the bipartite chunk--entity construction: entity linking creates a 2-hop path (chunk $\to$ entity $\to$ chunk) between any chunks sharing an entity mention (\S\ref{sec:results}). This is a structural invariant of bipartite entity-link graphs, independent of corpus size, NER quality, or document domain. Knowledge graphs with richer entity-to-entity relationships may exhibit leakage at PD $> 2$.

\subsection{The Amplification Mechanics}

The core insight is the \emph{identification of a compound attack surface} from composing two individually secure retrieval modalities. Vector retrieval with tenant prefiltering is secure ($\text{RPR} = 0.0$); graph retrieval within a single tenant's subgraph is secure. Connecting them---the pivot boundary---creates $70$--$194\times$ amplification ($\text{AF}(\epsilon)$) that neither modality exhibits alone, analogous to composition vulnerabilities in other security domains. The defense implication: authorization must be placed \emph{at the boundary}, after graph expansion and before context assembly. Authorization at the vector layer alone is insufficient; authorization at the LLM layer is unreliable.

\subsection{Implications for Agentic Systems}

The pivot attack is especially dangerous in agentic RAG deployments
(LangGraph, CrewAI) where graph traversal is performed autonomously
by an LLM agent~\cite{multiagentic2025}. In these systems, the
agent decides traversal depth, edge types, and expansion strategies
without human oversight. A poisoned vector seed can manipulate the
agent into performing unrestricted graph exploration, and research
shows that a single compromised agent can poison 87\% of downstream
decision-making within 4 hours~\cite{multiagentic2025}. Per-hop
authorization is even more critical in agentic settings, where there
is no human in the loop at the expansion step.

\section{Limitations}
\label{sec:limitations}

\textbf{Three corpora.} We evaluate on a synthetic enterprise corpus,
the Enron email corpus, and SEC EDGAR 10-K filings. The synthetic
corpus validates the vulnerability under controlled conditions; the
Enron and EDGAR corpora confirm generalization to real-world data
with varying entity connectivity densities. However, all three use a
bipartite chunk--entity graph topology that produces a uniform PD=2
signature.
Knowledge graphs with richer entity-to-entity relationships
(ontological hierarchies, multi-hop inference chains) may exhibit
leakage at deeper pivot depths. We do not evaluate on production
enterprise graphs with RBAC/ABAC policies, which may exhibit
different connectivity patterns.

\textbf{Generation evaluation scope.} Our generation contamination
experiment (\S\ref{sec:generation-impact}) evaluates three LLMs across
all three corpora ($n{=}40$--181 leaking queries on Enron, $n{=}64$--134 on
synthetic, $n{=}53$--139 on EDGAR) and confirms that leaked context
contaminates generated answers (ECR up to 0.32, FCR up to 0.07). The
GPT-4o synthetic sample ($n{=}10$) predates our expanded query pool;
larger samples across Enron and EDGAR provide more robust estimates.
We do not evaluate chain-of-thought reasoning, which might amplify the
impact by synthesizing cross-tenant connections.

\textbf{Two embedding models.} We evaluate two open-source models
(all-MiniLM-L6-v2, 384d, and all-mpnet-base-v2, 768d) and confirm
the vulnerability persists across both (\S\ref{sec:embedding}).
However, we do not test commercial embedding models (e.g., OpenAI
text-embedding-3-large) which may produce different similarity
distributions. The pivot vulnerability is structural (entity linking,
not embedding quality), so we expect model independence to hold broadly.

\textbf{Entity over-filtering.} D1's entity filtering trade-off and a finer-grained entity-aware authorization design are discussed in \S\ref{sec:mitigations}.

\textbf{Adaptive attacker scope.} We evaluate three adaptive attacks
(A5--A7) including targeted metadata forgery at rates up to 10\%,
entity manipulation, and query manipulation
(\S\ref{sec:adaptive-eval}). All defended pipelines maintain
$\text{RPR} = 0.0$. However, we do not evaluate attackers who craft
high-provenance payloads to bypass D4 specifically, or adversarial
query phrasing to manipulate D2's query classifier. Our metadata
forgery assumes the attacker can relabel node properties but cannot
modify the graph schema or index structure.

\textbf{NER and entity linker dependence.} The pivot path exists
because entity linking creates shared nodes across tenant boundaries.
All three corpora use spaCy's \texttt{en\_core\_web\_sm} NER model,
which has known recall limitations for domain-specific entities. On
the Enron corpus, this produces noisier entity extractions (dates,
monetary amounts, partial names) that create spurious cross-tenant
connections---yet RPR is lower (0.695 vs.\ 0.954), suggesting that
NER noise creates fragmented rather than dense pivot paths.
A higher-recall linker would extract more shared entities, potentially
increasing organic leakage. We do not evaluate how linker quality
affects RPR, nor do we test production entity linking systems that
perform coreference resolution or cross-document entity merging.

\textbf{Simplified policy model.} Our authorization model uses
tenant labels and four sensitivity tiers (PUBLIC through RESTRICTED).
Production enterprises typically employ richer access control:
role-based (RBAC) and attribute-based (ABAC) policies, group
memberships, temporary grants, legal holds, and need-to-know
exceptions. D1's per-hop check generalizes to any predicate that can
be evaluated on node metadata, but we do not demonstrate this
generality experimentally. The gap between our flat tenant model and
real-world RBAC/ABAC hierarchies remains an open integration question.

\section{Related Work}
\label{sec:related_work}

\subsection{Vector RAG Poisoning}

PoisonedRAG~\cite{zou2025poisonedrag} formalized knowledge corruption
attacks on RAG, achieving 90\% ASR with 5 injected documents.
CorruptRAG~\cite{corruptrag2025} demonstrated single-document attacks
with higher stealth. CtrlRAG~\cite{ctrlrag2025} achieved 90\% ASR on
GPT-4o via black-box feedback optimization.
RIPRAG~\cite{riprag2025} applied reinforcement learning to optimize
poisoning without model access.
NeuroGenPoisoning~\cite{neurogenpoison2025} targeted specific neurons
for >90\% success. These works study vector-side attacks in isolation;
none address what happens when poisoned vector results seed graph
expansion.

\subsection{GraphRAG Security}

GRAGPoison~\cite{gragpoison2026} is the closest related work,
demonstrating relation-centric poisoning of GraphRAG with 98\% ASR.
TKPA~\cite{fewwords2025} showed that modifying 0.06\% of corpus text
drops GraphRAG accuracy by 45 percentage points.
RAGCrawler~\cite{ragcrawler2026} achieved 84.4\% knowledge extraction
coverage through graph-guided probing. The Graph RAG Privacy
Paradox~\cite{graphragprivacy2025} established that graph RAG
increases structural leakage while reducing text leakage. Our work
extends these graph-side insights to the hybrid setting, where the
vector-to-graph transition creates a compound attack surface.

\subsection{RAG Privacy and Extraction}

The SoK on RAG Privacy~\cite{sokprivacy2026} systematized all known
privacy attack vectors in RAG systems and explicitly noted that hybrid
RAG privacy risks remain under-studied. Riddle Me
This~\cite{riddleme2025} demonstrated membership inference on RAG
systems. Traceback of RAG
Poisoning~\cite{traceback2025} provided forensic methods for
identifying responsible documents, but traces only through the vector
retrieval path without following graph expansion chains.

\subsection{RAG Defenses}

RAGuard~\cite{raguard2025} detects poisoning via retrieval pattern
analysis. SeCon-RAG~\cite{seconrag2025} applies semantic consistency
filtering. RevPRAG~\cite{revprag2025} achieves 98\% detection via
LLM activation analysis. SDAG~\cite{sdag2026} partitions context
into trusted and untrusted segments with block-sparse attention.
SD-RAG~\cite{sdrag2026} implements selective disclosure policies.
All operate within a single retrieval modality. Our defense suite
(D1--D5) is the first designed specifically for the cross-store
boundary, with per-hop authorization as the cornerstone mechanism.

\section*{Ethical Considerations}

Experiments use three corpora: a synthetic corpus generated by the
authors, the Enron email corpus (a public-record dataset
released during the 2001 FERC investigation and widely used in NLP
research~\cite{enroncorpus2015}), and SEC EDGAR 10-K filings (publicly
available regulatory documents). No production enterprise data, user
accounts, or confidential documents were used. The attack
implementations (A1--A7) target our own evaluation infrastructure
and are designed to characterize vulnerabilities, not exploit
production systems. Our primary contribution is the defense (D1),
which we release alongside the attack code to ensure that the net
effect of publication is protective. The released code does not
include tools for targeting external systems; all components require
a local Neo4j and ChromaDB deployment to operate.

\section*{Open Science}

We release the complete research artifact: source code, synthetic data
generators, Enron ingestion pipeline, all seven attack implementations
(A1--A7), the five-layer defense suite (D1--D5), evaluation harness,
query templates, and raw experimental results. The repository includes
255 passing unit tests and deterministic reproduction via a fixed random
seed (42). All experiments run on commodity hardware ($<$16\,GB RAM,
CPU-only) using open-source models and databases (spaCy, Sentence
Transformers, Neo4j Community Edition, ChromaDB). The Enron email corpus
is a public-record dataset available from Carnegie Mellon
University~\cite{enroncorpus2015}. No proprietary models, commercial APIs,
or restricted datasets are required to reproduce the core retrieval pivot
results. The generation contamination evaluation
(\S\ref{sec:generation-impact}) uses commercial LLM APIs (OpenAI,
Anthropic, DeepSeek) and is therefore dependent on API availability and
pricing; we include cached results for reproducibility without API access.

\section{Conclusion}
\label{sec:conclusion}

Hybrid RAG pipelines that combine vector retrieval with knowledge
graph expansion introduce an attack surface at the vector-to-graph
boundary. We formalized this threat as Retrieval Pivot Risk (RPR) and
demonstrated across three corpora---synthetic (RPR\,$=$\,0.95), Enron
(RPR\,$=$\,0.70), and EDGAR (RPR\,$=$\,0.09)---that undefended hybrid
pipelines exhibit high leakage rates compared to $\text{RPR} = 0.0$ for
vector-only retrieval, even without adversarial injection.

All leakage occurs at exactly $\text{PD} = 2$ hops---a structural
invariant of bipartite chunk--entity graphs that holds across all
three corpora regardless of size (2.8K--376K nodes), NER quality, or
document domain.

The most important finding is practical: \textbf{per-hop
authorization (D1)---re-checking tenant and sensitivity labels at
each graph expansion step---eliminates all measured leakage across
all three corpora, all seven attack variants, and metadata forgery
rates up to 10\%.} D1 requires no model changes, adds $<$1ms
latency, uses metadata already present in graph databases, and
retains 5.6$\times$ more context than vector-only retrieval. We
recommend it as the minimum viable security control for any hybrid
RAG deployment.

\bibliographystyle{plain}

\appendix

\section{Connectivity Sensitivity}
\label{app:connectivity}

To measure how shared-entity density affects leakage, we regenerate
the corpus with bridge entity counts $\in \{0, 5, 10, 15, 25, 40\}$,
rebuild all indexes for each count, and run 100 adversarial queries
through P3 (Table~\ref{tab:connectivity}).

\begin{table}[h]
\centering
\caption{Connectivity sweep: effect of bridge entity count on P3
  RPR and mean Leakage@k.}
\label{tab:connectivity}
\small
\begin{tabular}{r|ccc}
\toprule
\textbf{Bridges} & \textbf{P3 RPR} & \textbf{Mean Leak} &
  \textbf{PD} \\
\midrule
  0  & 0.93 & 21.5 & 2.0 \\
  5  & 0.95 & 26.6 & 2.0 \\
  10 & 0.95 & 30.5 & 2.0 \\
  15 & 0.95 & 31.5 & 2.0 \\
  25 & 0.95 & 32.5 & 2.0 \\
  40 & 0.94 & 34.5 & 2.0 \\
\bottomrule
\end{tabular}
\end{table}

Two findings emerge. First, \textbf{RPR is remarkably stable
(0.93--0.95) regardless of bridge count}---even with zero intentional
bridge entities, organic entity overlap from spaCy NER produces
RPR~$= 0.93$. Second, \textbf{mean leakage scales monotonically}
from 21.5 (0 bridges) to 34.5 (40 bridges), a 60\% increase. Bridge
entities do not \emph{enable} leakage (BFS already reaches
cross-tenant nodes through generic NER entities), but they
\emph{amplify} the volume of leaked content per query. PD remains
uniformly 2.0 across all bridge counts, confirming the structural
bipartite pivot.

\section{Embedding Model Sensitivity}
\label{app:embedding}

To verify that the pivot vulnerability is structural rather than
embedding-dependent, we repeat the P1 and P3 evaluations using
all-mpnet-base-v2 (768 dimensions, higher retrieval quality) in
addition to our default all-MiniLM-L6-v2 (384 dimensions). We rebuild
the ChromaDB collection for each model and run the full 500-query
evaluation (Table~\ref{tab:embedding-sensitivity}).

\begin{table}[h]
\centering
\caption{Embedding model sensitivity: P3 RPR and mean Leakage@k
  across two embedding models.}
\label{tab:embedding-sensitivity}
\small
\begin{tabular}{l|cc|cc}
\toprule
\textbf{Model} & \multicolumn{2}{c|}{\textbf{Benign}} &
  \multicolumn{2}{c}{\textbf{Adversarial}} \\
 & RPR & Leak & RPR & Leak \\
\midrule
  MiniLM-L6 (384d) & 0.954 & 16.0 & 0.947 & 19.4 \\
  MPNet (768d) & 0.994 & 19.4 & 0.980 & 28.4 \\
\bottomrule
\end{tabular}
\end{table}

The higher-quality MPNet model actually \emph{increases} RPR: from
0.954 to 0.994 on benign queries, and from 0.947 to 0.980 on
adversarial queries. Mean leakage also rises---from 16.0 to 19.4
(benign) and from 19.4 to 28.4 (adversarial). Better embeddings
retrieve more relevant seed chunks, which in turn mention more
entities, which seed more graph expansion into cross-tenant
neighborhoods. Both models produce $\text{PD} = 2.0$ uniformly,
confirming that the bipartite pivot structure is
model-independent. P1 achieves $\text{RPR} = 0.0$ under both
models, confirming that vector-side tenant prefiltering remains
effective regardless of embedding quality.

\section{Adaptive Attacks (A5--A7)}
\label{app:adaptive}

We extend the attack taxonomy with three adaptive strategies that
target the defense mechanisms themselves
(Table~\ref{tab:adaptive-attacks}).

\begin{table}[h]
\centering
\caption{RPR under adaptive attacks (A5--A7) across pipeline variants.
A5 tests metadata forgery at three rates; A6 tests entity manipulation;
A7 tests query manipulation. Leakage@k shown in parentheses.}
\label{tab:adaptive-attacks}
\small
\begin{tabular}{ll|cccc}
\toprule
\textbf{Attack} & \textbf{Rate} & \textbf{P3} & \textbf{P4} & \textbf{P7} & \textbf{P8} \\
\midrule
  A5 & 1\%  & 1.00 (7.8) & 0.00 (0.0) & 0.00 (0.0) & 0.00 (0.0) \\
  A5 & 5\%  & 1.00 (7.8) & 0.00 (0.0) & 0.00 (0.0) & 0.00 (0.0) \\
  A5 & 10\% & 1.00 (7.8) & 0.00 (0.0) & 0.00 (0.0) & 0.00 (0.0) \\
\midrule
  A6 & ---  & 1.00 (7.8) & 0.00 (0.0) & 0.00 (0.0) & 0.00 (0.0) \\
  A7 & ---  & \multicolumn{4}{c}{\textit{query-only (no injection)}} \\
\bottomrule
\end{tabular}
\end{table}

\textbf{A5 (Metadata Forgery)} relabels injected nodes with the
target tenant's name to bypass D1's tenant check. At forgery rates of
1\%, 5\%, and 10\%, A5 achieves $\text{RPR} = 1.0$ against P3 but
$\text{RPR} = 0.0$ against all defended pipelines (P4, P7, P8).
D1 remains effective because the forged metadata only affects the
attacker's own injected nodes---the organic entity nodes that create
the pivot path still carry empty-string tenant labels and are filtered.

\textbf{A6 (Entity Manipulation)} creates documents mentioning
entities from the target tenant's namespace, attempting to create new
shared entity nodes. A6 also fails against D1: the newly created
entity nodes still carry empty-string tenant labels, and the defense
filters them regardless of how they were created.

\textbf{A7 (Query Manipulation)} crafts queries mentioning target
tenant entity names to steer NER-based entity linking toward sensitive
neighborhoods. This is a query-only attack (no injection) and
produces the same RPR as organic benign queries---confirming that the
vulnerability is in the graph expansion, not in query processing.

\section{Generation Impact}
\label{app:generation}

To measure whether leaked context actually contaminates LLM-generated
answers, we evaluate production LLMs on queries with known leakage
across all three corpora (Table~\ref{tab:generation}).

\begin{table}[h]
\centering
\caption{Generation contamination metrics across datasets and LLMs.
ECR = Entity Contamination Rate, ILS = Information Leakage Score,
FCR = Factual Contamination Rate, GRR = Generation Refusal Rate.
$n$ = queries with leakage evaluated.}
\label{tab:generation}
\small
\begin{tabular}{ll|cccc|r}
\toprule
\textbf{Dataset} & \textbf{LLM} & \textbf{ECR} & \textbf{ILS} & \textbf{FCR} & \textbf{GRR} & $n$ \\
\midrule
\multirow{3}{*}{Synthetic}
  & GPT-4o            & 0.077 & 0.305 & 0.050 & 0.800 & 10 \\
  & Claude Sonnet 4.5 & 0.321 & 0.352 & 0.072 & 0.047 & 64 \\
  & DeepSeek-V3       & 0.147 & 0.333 & 0.044 & 0.045 & 134 \\
\midrule
\multirow{3}{*}{Enron}
  & GPT-5.2           & 0.082 & 0.285 & 0.005 & 0.000 & 181 \\
  & Claude Sonnet 4.5 & 0.133 & 0.280 & 0.004 & 0.000 & 40 \\
  & DeepSeek-V3       & 0.053 & 0.260 & 0.056 & 0.050 & 101 \\
\midrule
\multirow{2}{*}{EDGAR}
  & Claude Sonnet 4.5 & 0.040 & 0.358 & 0.000 & 0.019 & 53 \\
  & DeepSeek-V3       & 0.025 & 0.345 & 0.016 & 0.101 & 139 \\
\bottomrule
\end{tabular}
\vspace{2pt}
\begin{minipage}{0.95\linewidth}
\footnotesize
\textit{Note:} ECR drops from synthetic to Enron to EDGAR as entity
distinctiveness decreases.
\end{minipage}
\end{table}

The results reveal model-dependent and dataset-dependent contamination
behavior. On the synthetic corpus, Claude Sonnet 4.5 exhibits the
highest Entity Contamination Rate ($\text{ECR} = 0.32$, $n{=}64$),
indicating it readily incorporates leaked entities into generated
answers. GPT-4o shows the opposite pattern: $\text{ECR} = 0.08$ and
$\text{GRR} = 0.80$ ($n{=}10$), largely ignoring leaked context.
DeepSeek-V3 falls between ($\text{ECR} = 0.15$, $n{=}134$) with
low GRR ($0.05$) and FCR ($0.04$). All three models show GRR
$\leq 0.05$ at scale, suggesting that generation refusal is rare
when leaked context is topically relevant.

On the Enron corpus ($n{=}181$ for GPT-5.2, $n{=}40$ for Claude,
$n{=}101$ for DeepSeek), ECR drops substantially across all models
(GPT-5.2 $0.08$; Claude $0.13$; DeepSeek $0.05$) while GRR
approaches zero (all models $\leq 0.05$). Real email content is
contextually relevant enough that LLMs rarely ignore it, but leaked
entities from adjacent departments are less distinctive than synthetic
ones, producing lower entity contamination. GPT-5.2 shows moderate
ECR ($0.082$) with near-zero FCR ($0.005$), suggesting it surfaces
entity names but avoids reproducing substantive facts from leaked
context. These findings confirm that retrieval-level leakage
translates to generation-level contamination across all three corpora, that
the severity depends on the model and the domain, and that the effect
is robust across sample sizes ($n = 40$--$181$).

\section{Traversal Parameter Sweep}
\label{app:traversal}

\begin{figure}[h]
\centering
\includegraphics[width=\columnwidth]{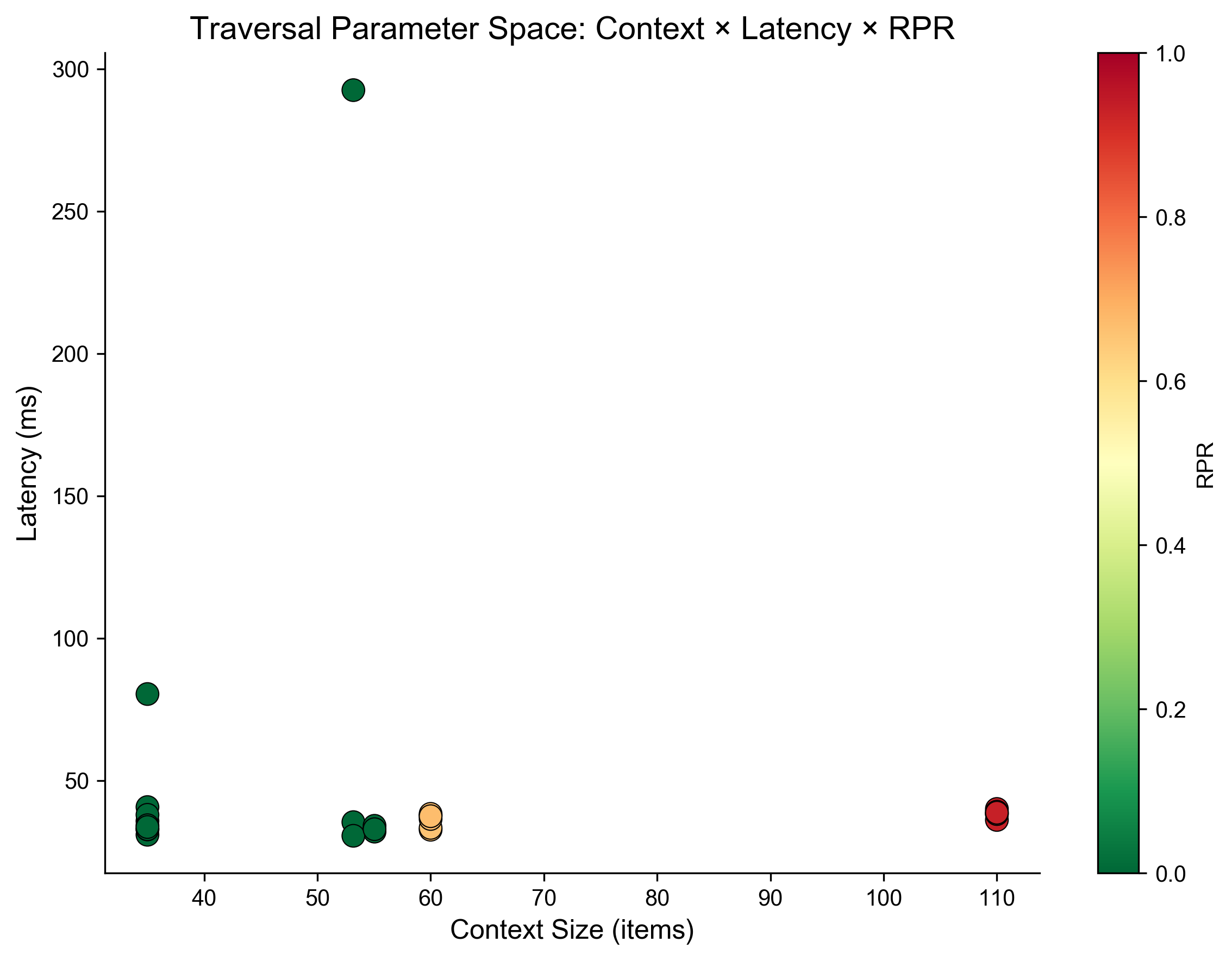}
\caption{Traversal parameter sweep: context size vs.~latency,
  colored by RPR. The total node budget is the primary
  leakage-controlling parameter.}
\label{fig:traversal}
\end{figure}

To understand which traversal parameters control leakage, we sweep
across 27 configurations: depth $\in \{1, 2, 3\}$, branching factor
$\in \{5, 10, 25\}$, and total node budget $\in \{25, 50, 100\}$,
running 100 adversarial queries per configuration against the
undefended hybrid pipeline (P3). Figure~\ref{fig:traversal} shows
the results.

Three findings emerge:

\textbf{Total node budget is the primary control.} Regardless of
depth or branching, $\text{total\_nodes} = 25$ yields $\text{RPR} = 0$
(insufficient expansion to reach cross-tenant content),
$\text{total\_nodes} = 50$ yields $\text{RPR} \approx 0.66$, and
$\text{total\_nodes} = 100$ yields $\text{RPR} \approx 0.93$. This
is because the total node budget caps how many graph nodes are
gathered, and once this cap prevents expansion from reaching the 2-hop
cross-tenant chunks, leakage is eliminated.

\textbf{Depth must be $\geq 2$ for leakage to occur.} At
$\text{depth} = 1$, $\text{RPR} = 0$ regardless of branching or
total node budget, because single-hop expansion cannot cross the
chunk$\to$entity$\to$chunk pivot path. This confirms the structural
PD=2 signature.

\textbf{Branching factor is irrelevant given a total node cap.} At
fixed depth and total node budget, branching $\in \{5, 10, 25\}$
produces identical RPR values. The BFS expansion fills the node budget
regardless of how many children are explored per node.

\section{Latency and Overhead}
\label{app:latency}

Table~\ref{tab:latency} presents latency and context size across
pipeline variants.

\begin{table}[h]
\centering
\caption{Latency (ms) and context size across pipeline variants.}
\label{tab:latency}
\small
\begin{tabular}{l|ccc|c}
\toprule
\textbf{Variant} & \textbf{p50} & \textbf{p95} & \textbf{Mean} &
  \textbf{Ctx} \\
\midrule
\multicolumn{5}{c}{\textit{Benign Queries}} \\
\midrule
  P1 (Vector) & 12.6 & 19.9 & 16.2 & 10 \\
  P3 (Hybrid) & 26.7 & 38.1 & 32.4 & 110 \\
  P4 (+D1) & 26.5 & 34.4 & 30.4 & 56 \\
  P6 (+D1--D3) & 23.1 & 29.8 & 26.4 & 29 \\
  P8 (All) & 22.7 & 28.6 & 25.6 & 20 \\
\midrule
\multicolumn{5}{c}{\textit{Adversarial Queries}} \\
\midrule
  P1 (Vector) & 12.3 & 34.6 & 23.5 & 10 \\
  P3 (Hybrid) & 27.3 & 40.2 & 33.7 & 110 \\
  P4 (+D1) & 26.3 & 32.2 & 29.3 & 50 \\
  P6 (+D1--D3) & 23.1 & 33.7 & 28.4 & 28 \\
  P8 (All) & 22.9 & 30.1 & 26.5 & 20 \\
\bottomrule
\end{tabular}
\end{table}

D1 adds negligible latency overhead---P3 to P4 shows $<$1ms increase
at p50 (26.7$\to$26.5ms benign). More striking, the full defense
stack (P8) is actually \emph{faster} than undefended P3 (22.7 vs.
26.7ms) because budgeted traversal (D3) and trust filtering (D4)
reduce the number of nodes expanded and processed, decreasing both
graph query time and context assembly overhead.

\section{Metadata Integrity Stress Test}
\label{app:mislabel}

An adaptive attacker might attempt to circumvent D1 by corrupting
sensitivity labels. We test D1's robustness by randomly flipping
sensitivity labels on $r\%$ of graph nodes
($r \in \{0.1, 0.5, 1.0, 2.0, 5.0\}$) and measuring RPR under
P4. Results: \textbf{D1 maintains $\text{RPR} = 0.0$ at all
mislabel rates up to 5\%.} This robustness arises because D1's
primary protection is the \emph{tenant} filter (not the sensitivity
filter): cross-tenant leakage requires traversing to a node with a
different tenant label, and sensitivity mislabeling does not affect
tenant assignments. We note that context size decreases slightly at
higher mislabel rates (50$\to$48 items at 5\%) as some authorized
nodes have their sensitivity erroneously raised above the user's
clearance.

\begin{figure}[h]
\centering
\includegraphics[width=\columnwidth]{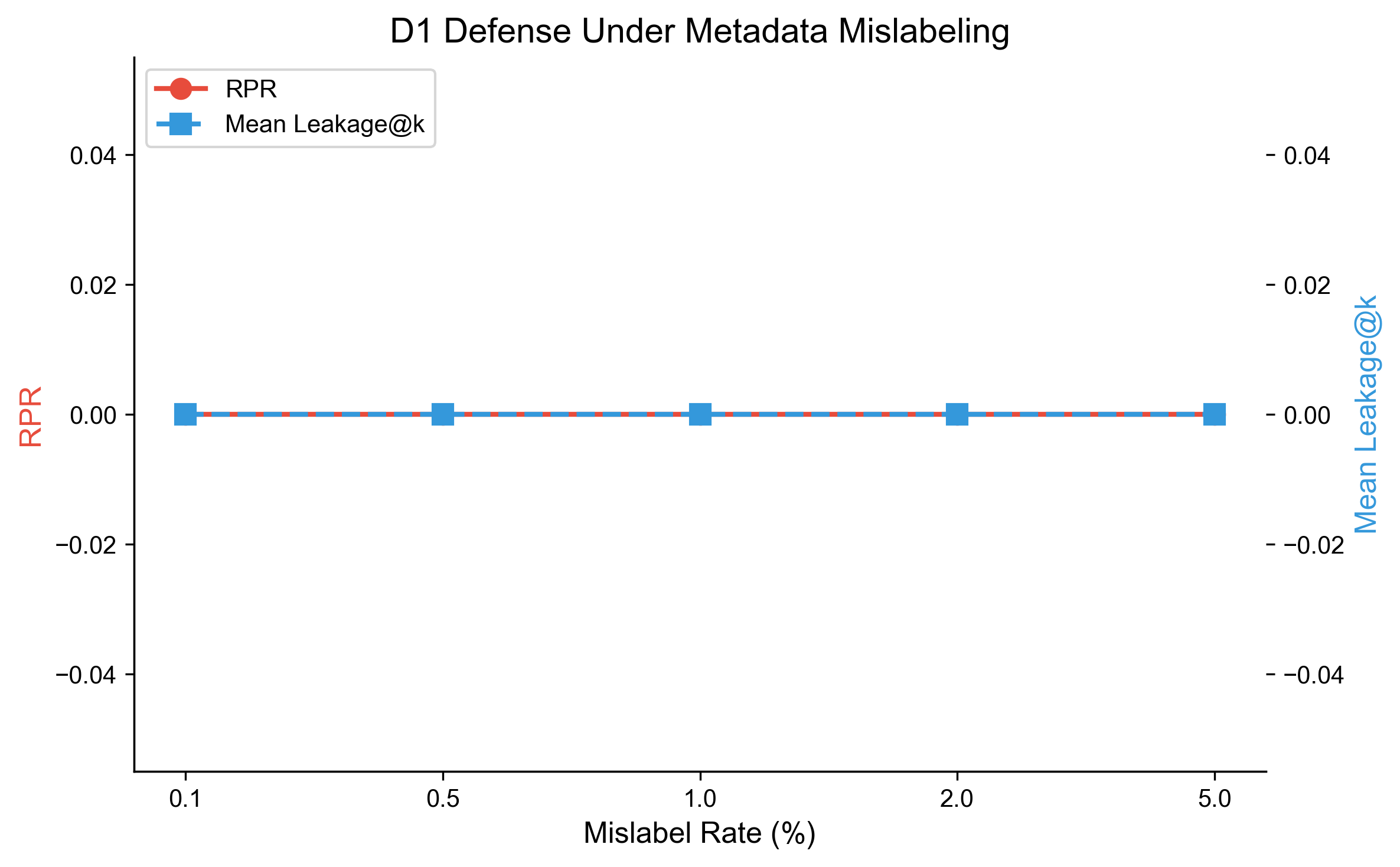}
\caption{D1 robustness under metadata corruption. RPR remains 0.0
  even at 5\% mislabel rate. Context size decreases slightly as
  erroneously up-labeled nodes are filtered.}
\label{fig:mislabel}
\end{figure}

\section{Utility Impact}
\label{app:utility}

A critical question for practitioners is whether defenses degrade
retrieval quality. Since our evaluation corpus uses synthetic documents
without human-annotated relevance labels, we measure utility through two
proxy metrics: (1)~\textbf{authorized context size}---the number of
items passing authorization checks in the final context, and
(2)~\textbf{authorization rate}---the fraction of context items that are
authorized (i.e., not leaked). Together these quantify how much useful
content each defense preserves and how much noise it eliminates.

Table~\ref{tab:utility} presents the security-utility tradeoff. P3
(undefended hybrid) returns 110 items per query, but only 85.5\% are
authorized---the remaining 16 items (14.5\%) are cross-tenant or
over-clearance leakage. P1 (vector-only) returns 10 fully authorized
items, establishing the vector-only utility baseline.

\begin{table}[h]
\centering
\caption{Security-utility tradeoff across pipeline variants
  (benign queries, $n{=}350$). Auth.\ Items = mean authorized items in
  context. Auth.\ Rate = fraction of context that is authorized.
  Retention = authorized items relative to P3's authorized baseline
  (94 items).}
\label{tab:utility}
\small
\begin{tabular}{l|ccccc}
\toprule
\textbf{Variant} & \textbf{RPR} & \textbf{Ctx}
  & \textbf{Auth.} & \textbf{Auth.\ Rate} & \textbf{Retention} \\
\midrule
P1 (vector) & 0.000 & 10 & 10.0 & 1.00 & --- \\
P3 (hybrid) & 0.954 & 110 & 94.0 & 0.85 & 1.00 \\
P4 (+D1) & 0.000 & 56 & 56.0 & 1.00 & 0.60 \\
P6 (+D1--3) & 0.000 & 29 & 29.0 & 1.00 & 0.31 \\
P8 (+D1--5) & 0.000 & 20 & 20.0 & 1.00 & 0.21 \\
\bottomrule
\end{tabular}
\end{table}

D1 (P4) retains 56 authorized items per query---a 40\% reduction from
P3's 94 authorized items, driven entirely by entity node removal
(\S\ref{sec:mitigations}). However, P4 still provides
5.6$\times$ more authorized content than P1's vector-only baseline (56
vs.\ 10), confirming that \emph{graph expansion retains substantial
utility even after D1 filtering}. The lost items are entity nodes and
their dependent traversal paths, not authorized chunks from the user's
own tenant.

D3--D5 progressively reduce context size (56 $\to$ 29 $\to$ 20) by
capping traversal depth, filtering low-provenance nodes, and removing
noise at the merge stage. Even the most aggressive configuration (P8)
retains 2$\times$ more content than vector-only retrieval. Under
adversarial queries, the pattern is similar: P4 retains 50 authorized
items (retention 0.55 vs.\ P3's 90.6 authorized baseline).

The practical conclusion: \textbf{D1 eliminates all leakage while
preserving 5.6$\times$ the content of vector-only retrieval.} The
additional defenses (D3--D5) are context quality optimizers that reduce
noise for downstream LLM generation. Organizations prioritizing breadth
should deploy D1 alone; those prioritizing signal-to-noise ratio should
add D3 and D4.

\section{Severity-Weighted Leakage}
\label{app:swl}

Not all leakage is equally harmful. A PUBLIC-clearance user receiving
INTERNAL content is a lesser violation than receiving RESTRICTED
content. We define severity-weighted leakage as:

\begin{equation}
\text{SWL}(q, u) = \sum_{\substack{x \in S_k(q) \\ \text{Sensitive}(x,u)}}
  w(x, u)
\label{eq:swl}
\end{equation}

\noindent where the per-item weight $w(x, u)$ captures two distinct
policy violations:

\begin{equation}
w(x, u) = \begin{cases}
  \ell_x - \ell_u & \text{if } \ell_x > \ell_u \\
  1 & \text{if } t_x \neq t_u \wedge \ell_x \leq \ell_u
\end{cases}
\label{eq:swl_weight}
\end{equation}

\noindent Here $\ell_x$ is the sensitivity level (PUBLIC=0,
INTERNAL=1, CONFIDENTIAL=2, RESTRICTED=3) and $t_x$ the tenant of
item $x$. Over-clearance items contribute weight proportional to the
sensitivity gap. Cross-tenant items that do not exceed the user's
clearance receive a penalty of 1, reflecting the policy violation of
accessing another tenant's data regardless of sensitivity.

\section{Artifact Appendix}
\label{sec:artifact}

\subsection{Repository}

The complete codebase, synthetic data generators, attack
implementations, defense suite, evaluation harness, and experimental
results are available at:
\url{https://github.com/scthornton/hybrid-rag-pivot-attacks}

\subsection{System Requirements}

\begin{itemize}
\item Python 3.11+ with dependencies: \texttt{chromadb},
  \texttt{neo4j}, \texttt{spacy}, \texttt{sentence-transformers},
  \texttt{pydantic}, \texttt{numpy}
\item Neo4j 5.15+ with APOC plugin (for
  \texttt{apoc.path.spanningTree})
\item ChromaDB server (latest)
\item spaCy model: \texttt{en\_core\_web\_sm}
\end{itemize}

\subsection{Reproduction Steps}

\textbf{Synthetic corpus:}
\begin{enumerate}
\item Clone repository and install: \texttt{pip install -e ".[dev]"}
\item Start services: \texttt{docker compose up -d} (Neo4j + ChromaDB)
\item Generate corpus: \texttt{python scripts/make\_synth\_data.py}
\item Generate queries: \texttt{python scripts/generate\_queries.py}
\item Build indexes: \texttt{python scripts/build\_indexes.py}
\item Run experiments: \texttt{python scripts/run\_experiments.py
  --bootstrap}
\item Run attack experiments: \texttt{python
  scripts/run\_attack\_experiments.py}
\item Run sweeps: \texttt{python scripts/run\_sweep\_experiments.py
  --traversal-sweep --mislabel-sweep --connectivity-sweep}
\end{enumerate}

\textbf{Enron corpus:}
\begin{enumerate}
\item Ingest Enron emails: \texttt{python scripts/ingest\_enron.py}
\item Build indexes: \texttt{python scripts/build\_indexes.py
  --dataset enron}
\item Run experiments: \texttt{python scripts/run\_experiments.py
  --dataset enron}
\end{enumerate}

\noindent Run tests: \texttt{pytest tests/ -v} (255 tests passing).

\subsection{Runtime Estimates}

\textbf{Synthetic:} Corpus generation and index building complete in
$\sim$5 minutes.  The full 500-query evaluation across 7 pipeline
variants takes $\sim$20 minutes.  Attack experiments
($4 \times 4$ configurations) require $\sim$10 minutes including graph
rebuilds. Sweep experiments total $\sim$45 minutes.

\textbf{Enron:} Ingestion and NER extraction takes $\sim$40 minutes
(2.07M entity mentions across 152K chunks). Neo4j graph loading
completes in $\sim$2 minutes using batch UNWIND queries. The 200-query
evaluation across P1/P3/P4 takes $\sim$10 minutes.

All experiments require $<$16\,GB RAM and run on CPU-only hardware.

\section{Corpus Generator Details}
\label{sec:corpus-details}

The synthetic corpus is produced by 12 domain-specific document
generators (3 per tenant), each instantiating one of three genre
templates: \emph{architecture description}, \emph{policy/procedure},
and \emph{status report}. Each generator populates templates from
curated entity pools:

\begin{itemize}
\item \textbf{Engineering:} 12 system names, 15 technologies, 6
  projects (e.g., \texttt{auth-service}, \texttt{Kubernetes},
  \texttt{Project Alpha}).
\item \textbf{Finance:} 10 vendors, 6 accounts, 7 regulations
  (e.g., \texttt{Deloitte}, \texttt{SOX}, \texttt{Capital Expenditure
  2025}).
\item \textbf{HR:} 12 departments, 7 benefits, 10 named employees
  (e.g., \texttt{Engineering}, \texttt{401k matching}, \texttt{Maria
  Chen}).
\item \textbf{Security:} 6 CVEs, 8 tools, 6 frameworks
  (e.g., \texttt{CVE-2025-41923}, \texttt{Splunk SIEM}, \texttt{NIST
  CSF}).
\end{itemize}

\noindent Bridge entities span 5 categories: shared vendors
(\texttt{CloudCorp}, \texttt{DataSyncInc}, \texttt{SecureNetLLC}),
shared infrastructure (\texttt{k8s-prod-cluster}, \texttt{splunk-siem},
\texttt{auth-service}), shared personnel (\texttt{Maria Chen},
\texttt{James Rodriguez}, \texttt{Aisha Patel}), shared compliance
(\texttt{SOC2-audit}, \texttt{PCI-DSS-cert}, \texttt{ISO27001}),
and shared projects (\texttt{ProjectNexus}, \texttt{ProjectHorizon},
\texttt{ProjectArcade}). Sensitivity tiers are assigned per-document:
PUBLIC 40\%, INTERNAL 30\%, CONFIDENTIAL 20\%, RESTRICTED 10\%.
All generation uses seed 42 for reproducibility.

\section{Query Templates}
\label{sec:query-templates}

\textbf{Benign queries} (350 total, 5 templates $\times$ 4 domains
$\times$ variable entity fills). Example templates:

\begin{small}
\begin{verbatim}
Engineering: "What is the architecture of {system}?"
Finance:     "Summarize the contract with {vendor}."
HR:          "What are the policies for {department}?"
Security:    "What is the severity of {cve}?"
\end{verbatim}
\end{small}

\noindent \textbf{Adversarial queries} (150 total) are stratified
across 4 attack types:

\begin{small}
\begin{verbatim}
A1 (Seed Steering):     "What systems like {system}
     handle {other_tenant_technology}?"
A2 (Entity Anchor):     "Tell me about {bridge_entity}
     and its role in {target_tenant_domain}."
A3 (Neighborhood Flood): "List everything related to
     {bridge_entity}."
A4 (Bridge Node):       "What connections exist between
     {bridge_entity} across departments?"
\end{verbatim}
\end{small}

\noindent Template slots are filled from aligned entity pools
(identical to those in the corpus generator) to ensure query
entities appear in the knowledge graph. Each query carries
metadata: \texttt{user\_tenant} (always
\texttt{acme\_engineering}), \texttt{user\_clearance} (stratified
across PUBLIC, INTERNAL, CONFIDENTIAL), and \texttt{query\_type}
(benign or adversarial with attack subtype).

\section{Defense Implementation}
\label{sec:defense-impl}

\textbf{D1 (Per-hop authorization).} After BFS expansion via
\texttt{apoc.path.spanningTree}, each returned node is checked:

\begin{small}
\begin{verbatim}
def is_node_authorized(node, user):
    tier = SensitivityTier(node.sensitivity)
    if tier > user.clearance:
        return False
    if node.tenant not in user.allowed_tenants:
        return False
    return True
\end{verbatim}
\end{small}

\noindent Entity nodes carry \texttt{tenant=""} (empty string),
so they always fail the tenant check. This is the mechanism that
severs the chunk$\to$entity$\to$chunk pivot path.

\textbf{D3 (Budgeted traversal).} The Cypher query enforces a global
node budget via \texttt{LIMIT \$max\_total}. Per-hop branching is
enforced post-query: expanded nodes are grouped by hop depth, and each
group is truncated to \texttt{max\_branching} entries.

\textbf{BFS Cypher query} (simplified):
\begin{small}
\begin{verbatim}
UNWIND $seed_ids AS seed_id
MATCH (start {node_id: seed_id})
CALL apoc.path.spanningTree(start, {
    maxLevel: $max_hops,
    limit: $max_total
}) YIELD path
WITH last(nodes(path)) AS node,
     length(path) AS depth
RETURN node.node_id, node.tenant,
       node.sensitivity, min(depth)
ORDER BY hop_depth LIMIT $max_total
\end{verbatim}
\end{small}

\noindent The \texttt{length(path)} return value feeds the Pivot
Depth metric directly.

\section{Entity-Aware Authorization}
\label{sec:entity-aware-future}

D1 eliminates all entity nodes because they carry empty-string tenant
labels. This is effective but coarse: it removes potentially useful
entity context (names, types, relations) from the retrieval result.
A finer-grained scheme would distinguish \emph{traversal authorization}
from \emph{inclusion authorization}:

\begin{enumerate}
\item \textbf{Entity tenant-set labeling.} During graph construction,
  label each entity with the set of tenants whose documents mention it:
  \texttt{tenant\_set = \{acme, globex\}}. Shared entities retain
  multiple tenant labels.
\item \textbf{Traversal-through permission.} Allow BFS to traverse
  \emph{through} an entity node if the user's tenant is in the entity's
  \texttt{tenant\_set}. The entity itself may appear in context.
\item \textbf{Destination check.} At each chunk node reached via an
  entity, re-check the chunk's tenant and sensitivity against the
  user's policy. This preserves D1's security guarantee at the
  chunk level while recovering entity utility.
\end{enumerate}

\noindent This scheme would increase context size (recovering the
entity nodes D1 currently removes) without re-enabling the
chunk$\to$entity$\to$\emph{unauthorized chunk} pivot, because the
destination check still blocks cross-tenant chunks.  The key trade-off
is implementation complexity: entity \texttt{tenant\_set} must be
maintained as documents are added or removed, and the authorization
predicate becomes a set-membership check rather than a simple equality.

\section{Statistical Methodology}
\label{sec:stats-appendix}

\textbf{Bootstrap procedure.} All confidence intervals use the
non-parametric percentile bootstrap (10{,}000 resamples, seed 42).
For binary RPR indicators, the bootstrap distribution is constructed by
resampling the $n=500$ per-query binary outcomes $\{0, 1\}$ with
replacement and computing the mean for each resample.  The 95\% CI is
the $[2.5\%, 97.5\%]$ percentile interval of the bootstrap
distribution.

\textbf{Sample size justification.} With $n = 500$ queries, we achieve
power $> 0.80$ for detecting a 5 percentage-point difference in RPR
(from 0.0 to 0.05) at $\alpha = 0.05$, computed via the exact binomial
test.  For leakage means, the standard error of the bootstrap is
$\text{SE} < 0.5$ items, sufficient to distinguish mean differences of
$\geq 1$ item.

\textbf{Multiple comparisons.} We test 7 pipeline variants against 2
query types (14 comparisons per metric).  Applying a Bonferroni
correction ($\alpha_\text{adj} = 0.05/14 = 0.0036$), the core findings
(P1/P4--P8: RPR $= 0.0$; P3: RPR $> 0.90$) remain significant at
$p < 10^{-6}$.

\textbf{$\epsilon$ sensitivity.} The regularized amplification factor
$\text{AF}(\epsilon)$ uses $\epsilon = 0.1$.  We verify that
conclusions are robust across $\epsilon \in \{0.01, 0.05, 0.1, 0.5\}$:
$\text{AF}(\epsilon)$ ranges from 1{,}599 ($\epsilon{=}0.01$) to 32
($\epsilon{=}0.5$), all confirming $>$$30\times$ amplification.

\end{document}